\documentclass[prd,preprintnumbers,superscriptaddress,nofootinbib,showpacs]{revtex4-1}
\pdfoutput=1
\usepackage{hyperref}
\usepackage{pstricks}
\usepackage{color}
\usepackage{amssymb,amsmath,bm}
\usepackage{epsfig}
\usepackage{afterpage}
\usepackage{longtable}
\usepackage{latexsym, mathrsfs}
\usepackage{graphics}
\usepackage{url}
\usepackage[left=2cm,right=2cm]{geometry}
\usepackage{subfigure}
\newcommand{\be}{\begin{equation}}
\newcommand{\ee}{\end{equation}}
\newcommand{\bea}{\begin{eqnarray}}
\newcommand{\eea}{\end{eqnarray}}
\newcommand{\bec}{\begin{center}}
\newcommand{\eec}{\end{center}}
\newcommand{\nn}{\nonumber}
\newcommand{\dd}{\displaystyle}

\begin{document}

\preprint{BARI-TH/16-704}

\title{Role of nonlocal probes of thermalization \\ for a strongly interacting non-Abelian plasma}
\author{L.~Bellantuono}
\affiliation{Dipartimento di Fisica, Universit\`a degli Studi di Bari, Italy}
\affiliation{Istituto Nazionale di Fisica Nucleare, Sezione di Bari, Italy}
\author{P.~Colangelo}
\affiliation{Istituto Nazionale di Fisica Nucleare, Sezione di Bari, Italy}
\author{F.~De~Fazio}
\affiliation{Istituto Nazionale di Fisica Nucleare, Sezione di Bari, Italy}
\author{F.~Giannuzzi}
\affiliation{Dipartimento di Fisica, Universit\`a degli Studi di Bari, Italy}
\author{S.~Nicotri}
\affiliation{Istituto Nazionale di Fisica Nucleare, Sezione di Bari, Italy}

\begin{abstract}
The thermalization process of an out-of-equilibrium  boost-invariant strongly interacting non-Abelian plasma is investigated using a holographic method. Boundary sourcing, a distortion of the boundary metric, is employed to drive the system far from equilibrium. Thermalization is analyzed in the fully dynamical system through nonlocal probes: the equal-time two-point correlation function of large conformal dimension operators in the boundary theory, and Wilson loops of different shapes. A dependence  of the thermalization time on the size of the probes is found, which can be compared to the result of local observables: the onset of thermalization is first observed at short distances.
\end{abstract}

\vspace*{2cm}

\maketitle

\section{Introduction}
The relativistic heavy ion collisions realized at the Brookhaven RHIC and at the CERN LHC produce a plasma-like system whose properties are similar to the ones expected for the quark-gluon plasma (QGP).
A dense strongly interacting medium is created, with the relevant degrees of freedom not represented by the individual partons, but more appropriately described as a fluid \cite{Heinz:2004pj,Shuryak:2014zxa}.
Simulations aimed at reproducing the experimental observations for, e.g.,  the elliptic flow which is due to the pressure anisotropy, point to an almost perfect fluid behavior, with a small viscosity to entropy density ratio $\dd{\eta/s}\sim {\cal O}$(0.1) and a short time of ${\cal O}$(1~fm/c) to reach thermal equilibrium.
Perturbative QCD calculations predict larger values of $\dd{\eta/s}$: this leads one to conclude that the created plasma is a realization of a strongly coupled deconfined phase of QCD.

Understanding in detail the features observed in experiment, in particular investigating the system early time dynamics, how the out-of-equilibrium strongly interacting plasma evolves towards a thermalized state, are challenging tasks.
Conventional methods make use of the idea that, soon after the collisions, ensembles of strong, coherent, longitudinal color electric and color magnetic fields are produced, then the evolution of such field configurations is numerically studied (see, e.g., \cite{Ruggieri:2015yea} and references therein).

A completely different theoretical approach exploits the gauge/gravity, in particular the AdS/CFT, duality.
This relates a strongly coupled gauge theory defined in a $d$-dimensional Minkowski space and a classical gravity theory living in a ($d$+1)-dimensional asymptotically AdS space times a compact manifold, with the Minkowski space representing the boundary of the AdS one \cite{Maldacena:1997re,Witten:1998qj,Gubser:1998bc}\footnote{For an introduction to the AdS/CFT principles and applications see the book \cite{Ammon:2015wua}.}.
The holographic approach permits to study how a strongly coupled system, driven out-of-equilibrium through an external quench, reaches a thermalized state, with a determination, e.g., of the time needed for the equilibration process.

In the gauge/gravity duality framework, thermalization in the gauge theory corresponds to the formation of a black brane in the higher dimensional space.
For an expanding strongly coupled plasma displaying a perfect fluid behavior, the boundary stress-energy tensor $T_{\mu\nu}$ components obey the Bjorken's relations \cite{Bjorken:1982qr}.
According to the AdS/CFT correspondence dictionary, each operator in the gauge theory has a dual field in the gravity side, and the stress-energy tensor has the metric as a dual.
The dual of $T_{\mu \nu}$ fulfilling Bjorken's hydrodynamics is identified as a black brane metric with time-dependent horizon \cite{Janik:2005zt}.
Corrections to the perfect fluid behavior can be analyzed through a corresponding gravity dual.\footnote{An overview of the fluid/gravity correspondence can be found in \cite{Hubeny:2011hd}. }

The early-time dynamics and subsequent thermalization  have been investigated in recent years,  employing holographic techniques within different contexts
  \cite{Kovchegov:2007pq,AbajoArrastia:2010yt,Bhattacharyya:2009uu,Albash:2010mv,Aparicio:2011zy,Keranen:2011xs,Galante:2012pv,Baron:2012fv,Caceres:2012em,Basu:2011ft,Auzzi:2013pca,Caceres:2013dma,Liu:2013iza,Basu:2013soa,
Zeng:2013fsa,Caceres:2014pda,Liu:2013qca,Fischler:2013fba,Hubeny:2013dea,Alishahiha:2014cwa,Fonda:2014ula,Zhang:2015dia}.
In particular, a way to study thermalization of a strongly coupled plasma has been proposed in Refs.~\cite{Chesler:2008hg,Chesler:2009cy} (and reviewed in \cite{Chesler:2013lia}): a distortion of the boundary metric is implemented to mimic an effect driving the system out of equilibrium.
The dual metric is computed by solving Einstein's equations in the higher dimensional space, and imposing appropriate boundary conditions.
Using the holographic renormalization procedure \cite{deHaro:2000xn}, the components of the boundary stress-energy tensor are determined in terms of the coefficients of the near-boundary expansion of the gravity metric.
When the boundary distortion, the {\it quench}, is impulsive with finite time duration, one can determine the elapsed time for the stress-energy tensor components to reach the hydrodynamic form after the quench is switched off, accessing the so-called thermalization time.
Hence, the observables are the system energy density and the pressures, which are studied as time proceeds.

Other studies deal with systems   put out of equilibrium through initial conditions, while the boundary metric is unperturbed \cite{Heller:2012km,Heller:2013oxa}. 
In this case the late-time physics is not described by hydrodynamics, and thermalization is related to holographic isotropization, with the thermalization time defined as the time after which pressure anisotropy is small compared to the energy density.
Another way of introducing a quench is letting the source of some operator vary with time, as in \cite{Ishii:2015gia} and references therein.
In particular, in \cite{Ishii:2015gia} a confining gauge theory is considered, in which a dilaton, with time-dependent UV boundary condition, breaks the conformal invariance.
Thermalization is studied by computing the time evolution of the energy-momentum tensor and of the one-point function of the scalar operator.
These investigations  can  provide useful hints  for the heavy ion collision phenomenology   \cite{CasalderreySolana:2011us,Arefeva:2012a}.

Beside local quantities, as the components of $T_{\mu\nu}$, the holographic methods permit one to access nonlocal probes evolving in time as well, such as the two-point correlation function of boundary theory operators, and the expectation values of Wilson loops defined on the boundary \cite{Balasubramanian:2010ce}.
Their calculation requires the length of the geodesics in the bulk connecting the two boundary points in the correlation function, or the area of the extremal surface plugging in the bulk and having the Wilson loop as a contour at the boundary.
Hence, such observables get information from deep regions in the bulk, accessing the IR regime of the boundary field theory.

An example of the use of nonlocal probes has been worked out using a Vaidya metric which describes the collapse of a thin mass shell from the boundary into the bulk \cite{Balasubramanian:2010ce,Balasubramanian:2011ur}.
At late times this metric coincides with a black hole one, dual to the thermalized gauge theory on the boundary. Thermalization has been studied by comparing the nonlocal probes in the Vaidya geometry with those obtained for a black hole metric with a time-independent horizon corresponding to the equilibrium temperature:  
it was observed that the time for thermalization depends on the size of the probe in the boundary theory.

The finite chemical potential case has been  investigated in a Reissner-Nordstr\"om-AdS black hole Vaidya-type metric \cite{Galante:2012pv}.
Larger thermalization times than those observed in the case of vanishing charge were found for the two-point function and for the expectation value of rectangular Wilson loops.
Different realisations have also been analytically studied \cite{Pedraza:2014moa}.
In \cite{Buchel:2014gta} a quench was introduced in the boundary theory through an operator with time-dependent mass dual to a scalar field in AdS black hole (AdS/BH) geometry.
Given the backreaction of the scalar field on the metric, the perturbed background has been computed using pseudospectral methods while the geometry thermalizes to a static black-hole AdS space, with a temperature higher than the initial one.
Local (e.g. one-point correlation function of the operator dual to the scalar field) and nonlocal (two-point correlation functions and entanglement entropy) observables have been investigated, finding again that the thermalization time  can depend on the size of the probe.

The above examples of nonlocal probes do not have a direct connection with the study of thermalization meant as the onset of a hydrodynamic regime, which is instead the main purpose of the present study.
We investigate the thermalization of a system taken out of equilibrium, computing the time needed by nonlocal observables to start behaving hydrodynamically. The comparison between the results obtained by local and nonlocal observables discloses different characteristic times.
We evaluate nonlocal observables in the case of two representative models of quenches scrutinised in \cite{Bellantuono:2015hxa}, using the computed solution of the Einstein equations and hence considering the  full dynamical system. 
In particular, two-point correlation functions of large dimension boundary operators and the expectation value of an infinite rectangular strip and of a circular Wilson loop are determined, and their time dependence is used to investigate the relaxation towards the hydrodynamic regime.
The distance between the points at which the correlation function is evaluated, and the size of Wilson loops represent the new variables in terms of which   thermalization is studied.

The plan of the paper is the following. In Section \ref{sec2} we review the results obtained in \cite{Bellantuono:2015hxa} using local observables, the energy density, entropy density and pressures,
for two models of quenches.
In Section \ref{sec3} we provide the expressions for the two-point correlation functions and the two kinds of Wilson loops used in the calculation. The results for the nonlocal probes are presented in Section \ref{sec4}, and the conclusions  are collected in the last Section.

\section{Thermalization by boundary sourcing:  results from local probes}\label{sec2}

In \cite{Bellantuono:2015hxa} the thermalization of a boost-invariant non-Abelian plasma has been studied adopting the method of boundary sourcing to drive the system far from equilibrium \cite{Chesler:2008hg,Chesler:2009cy}.
The stress-energy tensor of the boundary theory $ T^\mu_\nu =\frac{N_c^2}{2 \pi^2}\, {diag}(-\epsilon, p_\perp,p_\perp, p_\parallel)$ is written in terms of the system energy density $\epsilon$, of the pressure $p_\perp$ along one of the two transverse directions (with respect, e.g., to the heavy ion collision axis) and of the pressure $p_\parallel$ in the longitudinal direction.\footnote{Throughout the paper, the energy density and pressures are referred to without considering the factor $\frac{N_c^2}{2 \pi^2}$.}
The boundary 4$d$ coordinates are denoted as $x^\mu=(x^0,x^1,x^2,x^3)$, with $x^3=x_\parallel$ direction identified with the collision axis along which the plasma expands. The investigated system has boost invariance along this axis, together with translational and $O(2)$ rotational invariances in the transverse plane $x_\perp=\{x^1,\,x^2 \}$.
The 4$d$ line element: $ds^2_4=-d\tau^2+dx_\perp^2+\tau^2 dy^2$ is expressed in terms of the proper time $\tau$ and of the spacetime rapidity $y$, defined through $x^0=\tau \cosh y$ and $x_\parallel=\tau \sinh y$.

The system is driven out of equilibrium by a quench on the boundary metric.
The quench, described by the profile $\gamma(\tau)$, modifies the line element:
\be
ds^2_4=-d\tau^2+e^{\gamma(\tau)} dx_\perp^2+\tau^2 e^{-2\gamma(\tau)} dy^2 \,\, , \label{metric4D}
\ee
leaving the spatial three-volume unchanged and respecting the translational and $O(2)$ symmetries in the transverse plane.

The 5$d$ spacetime on which the gravity dual is defined, having the metric \eqref{metric4D} as a boundary, is described using
Eddington-Finkelstein coordinates, with $r$ the radial coordinate. The 5$d$ metric is written as
\be
ds^{2}=-A(r,\tau)d\tau^{2}+\Sigma(r,\tau)^{2}e^{B(r,\tau)}d {x}_{\perp}^{2}+\Sigma(r,\tau)^{2}e^{-2B(r,\tau)}dy^{2}+2d\tau dr \,\,\,.
\label{metric5D}
\ee
The boundary corresponds to $r \to \infty$.
The metric functions $A$, $\Sigma$ and $B$ depend only on $r$ and $\tau$ due to the chosen symmetries.
They have been determined by solving 5$d$ Einstein equations with a negative cosmological constant, that can be cast in the form \cite{Chesler:2009cy}:
\bea
&&\Sigma ({\dot \Sigma})^\prime +2 \Sigma^\prime {\dot \Sigma}-2 \Sigma^2=0 \nn \\
&& \Sigma ({\dot B})^\prime+\frac{3}{2} \left(\Sigma^\prime {\dot B}+B^\prime {\dot \Sigma}\right)=0 \nn \\
&& A^{\prime \prime} +3 B^\prime {\dot B} -12 \frac{\Sigma^\prime {\dot \Sigma} }{\Sigma^2}+4=0\label{ein} \\
&& {\ddot \Sigma}+\frac{1}{2} \left( {\dot B} ^2 \Sigma -A^\prime {\dot \Sigma} \right) =0 \nn \\
&& \Sigma^{\prime \prime}+\frac{1}{2} B^{\prime 2} \Sigma =0 \,. \nn
\eea
In \eqref{ein}, for a generic function $\xi(r,\tau)$, the derivatives $\xi^\prime=\partial_r \xi$ and ${\dot \xi}= \partial_\tau \xi+\frac{1}{2} A \partial_r \xi$ denote directional derivatives along the infalling radial null geodesics and the outgoing radial null geodesics, respectively.
Two boundary conditions are imposed.
The first states that the metric (\ref{metric5D}) produces the 4$d$ metric Eq.~(\ref{metric4D}) for $r \to \infty$.
Moreover, at the initial time slice $\tau=\tau_i$ when the distortion of the boundary metric is switched on, one has to start from the AdS$_5$ bulk metric:
\be
ds^2= r^2 \left[ -d \tau^2+ d x_\perp^2 + \left( \tau+\frac{1}{r} \right)^2 dy^2 \right] + 2 dr d\tau \,\,\, . \label{AdS5}
\ee
To investigate whether and how thermalization depends on particular boundary sourcing,
several distortion profiles have been considered in \cite{Bellantuono:2015hxa}. They are characterized by a function $\gamma(\tau)$ representing quenches with different number, structures and intensities, generically written as
\be
\gamma(\tau) =w \left[ \tanh \left(\frac{\tau -\tau_0}{\eta} \right) \right]^7 \,+\sum_{j=1}^{N}\gamma_j(\tau,\tau_{0,j}) \label{profile}
\ee
with
\be
\gamma_j(\tau,\tau_{0,j}) = c_j f_j(\tau,\tau_{0,j})^6 e^{-1/f_j(\tau,\tau_{0,j})} \Theta\left(1-\frac{(\tau-\tau_{0,j})^2}{\Delta_j^2}\right)\,\, \label{profile1}
\ee
and
\be
f_j(\tau,\tau_{0,j})= 1- \frac{(\tau-\tau_{0,j})^2}{\Delta_j^2} \,\,\, . \label{profile2}
\ee
The set of parameters $w, \eta, \tau_0, \tau_{0,j}, c_j$ and $ \Delta_j$ specifies the different quench models: here we focus on the models ${\cal B}$ and ${\cal A}(2)$ studied in Ref.\cite{Bellantuono:2015hxa}.
Model ${\cal A}(2)$ represents two short pulses in the boundary metric:  the parameters are set to $w=0$, $N=2$, $c_1 = 1$, $\Delta_{1,2}=1$, $\tau_{0,1}=\frac{5 }{4}\Delta_1$, $c_2 = 2$, $\tau_{0,2}=\frac{9}{4} \Delta_2$. The quench ends at $\tau_f^{\cal A}=3.25$.
Model $\cal B$ represents a slow deformation plus a short pulse, and is obtained using $w=\frac{2}{5}$, $\eta=1.2$, $\tau_0=0.25$, $N=1$, $c_1 = 1$, $\Delta_1=1$, $\tau_{0,1}=4 \Delta_1$.
The pulse ends at $\tau_f^{\cal B}=5$, while the slow distortion continues with $\tau$ and approaches a constant value.
In both cases, the quench is switched on at $\tau_i=0.25$.
The profiles $\gamma(\tau)$ are depicted in Fig.~\ref{fig:lgeo}.

The metric functions $A(r,\tau)$, $\Sigma(r,\tau)$ and $B(r,\tau)$ in \eqref{metric5D} have been computed in \cite{Bellantuono:2015hxa} by solving the Einstein equations \eqref{ein} (considered as three dynamical and two constraint equations), with the conditions provided at the initial time slice and on the boundary.
The solutions have allowed us to determine several quantities of interest, in particular the thermalization time obtained comparing the boundary stress-energy tensor to the viscous hydrodynamics behavior.
It is worth recalling that homogeneity, boost invariance and invariance under rotations in the transverse plane imply that the various components of $T_{\mu}^{\nu}$ depend only on the proper time $\tau$ \cite{Bjorken:1982qr}.
Moreover, for a conserved and traceless $T_{\mu}^{\nu}$ the components depend on a single function $f(\tau)$, so that $T_{\mu}^{\nu}$ can be written as
\be
T_{\mu}^{\nu}=diag \left(-f(\tau),\,f(\tau)+\displaystyle{\frac{1}{2}}\tau f^\prime(\tau),\,f(\tau)+\displaystyle{\frac{1}{2}}\tau f^\prime(\tau),\, -f(\tau)-\tau f^\prime(\tau)\right) \,\,\,\ .
\ee
For a perfect fluid, the equation of state $\epsilon=3 p$ and the relation $p=p_\parallel=p_\perp$ fix the $\tau$ dependence: $\epsilon(\tau)=\displaystyle{\frac{const}{\tau^{4/3}}}$, which is modified if viscous effects are included \cite{Janik:2005zt}.
An effective temperature $T_{eff}(\tau)$ can be defined through the relation $\epsilon(\tau)= \displaystyle{\frac{3}{4}} \pi^4 T_{eff}(\tau)^4$, and for $T_{eff}(\tau)$ the subleading terms in the large-$\tau$ expansion can be computed in ${\cal N}=4$ SYM, with the result \cite{Heller:2007qt}:
\bea
T_{eff}(\tau)&=&\frac{\Lambda}{(\Lambda \tau)^{1/3}} \Bigg[ 1-\frac{1}{6 \pi (\Lambda \tau)^{2/3}}+\frac{-1+\log 2}{36 \pi^2 (\Lambda \tau)^{4/3} }
+\frac{-21+2\pi^2+51 \log 2 -24 (\log 2)^2}{1944 \pi^3 (\Lambda \tau)^2} \nn \\ &+& {\cal O}\left( \frac{1}{(\Lambda \tau)^{8/3}} \right )\Bigg] \,\,\, , \label{Teff1}
\eea
and $\Lambda$ a parameter.
This expression corresponds for the energy density $\epsilon$, for the longitudinal $p_\parallel$ and for the transverse $p_\perp$ pressures, to the large $\tau$ dependence given by
\bea
\epsilon&=& \frac{3 \pi^4 \Lambda^4}{4 (\Lambda \tau)^{4/3} }\left[ 1-\frac{2c_1}{ (\Lambda \tau)^{2/3}}+\frac{c_2}{ (\Lambda \tau)^{4/3}} + {\cal O}\left( \frac{1}{(\Lambda \tau)^2} \right )\right] \,\,\, , \label{hydroeps} \\
p_\parallel (\tau)&=&\frac{ \pi^4 \Lambda^4}{ 4(\Lambda \tau)^{4/3} } \left[ 1-\frac{6c_1}{ (\Lambda \tau)^{2/3}}+\frac{5c_2}{ (\Lambda \tau)^{4/3}} + {\cal O}\left( \frac{1}{(\Lambda \tau)^2} \right )\right] \,\,\, , \label{hydroppar} \\
p_\perp (\tau) &=& \frac{ \pi^4 \Lambda^4}{ 4(\Lambda \tau)^{4/3} } \left[ 1-\frac{c_2}{ (\Lambda \tau)^{4/3}} + {\cal O}\left( \frac{1}{(\Lambda \tau)^2} \right ) \right] \,\,\, ,\label{hydropperp}
\eea
with $c_1=\displaystyle{\frac{1}{3 \pi}}$ and $c_2=\displaystyle{\frac{1+2 \log{2}}{18 \pi^2}}$.
$\Lambda$ depends on the quench model: the values $\Lambda^{\cal B}=1.12$ and $\Lambda^{{\cal A}}=1.73$
have been obtained in \cite{Bellantuono:2015hxa}. The large $\tau$ dependence of the pressure ratio $\dd \frac{p_\parallel}{p_\perp}$ and anisotropy $\dd \frac{\Delta p}{\epsilon}=\frac{p_\perp-p_\parallel}{\epsilon}$ derives from the above expressions.

The results  in \cite{Bellantuono:2015hxa} are obtained  comparing the energy density and pressures,  computed by the holographic renormalization procedure  from the explicit metric functions
$A(r,\tau)$, $\Sigma(r,\tau)$ and $B(r,\tau)$ in \eqref{metric5D}, with the asymptotic expressions \eqref{hydroeps}, \eqref{hydroppar} and \eqref{hydropperp}. The results can be summarized as follows.
Regardless of the quench, the energy density evolves according to the viscous hydrodynamic expression (\ref{hydroeps}) as soon as the impulsive quench is switched off, i.e. at $\tau_f^{\cal B}=5$ and $\tau_f^{\cal A}=3.25$ for two models of interest.
For pressures, a thermalization time $\tau_p$ can be defined, considering the system thermalized when the pressure ratio differs from the asymptotic expression obtained from \eqref{hydroppar} and \eqref{hydropperp} by less than 5$\%$.
In model ${\cal B}$ the thermalization time is $\tau_p^{\cal B}=6.74$, with a delay $\tau_p^{\cal B}-\tau_f^{\cal B}=1.74$; in model ${\cal A}(2)$ the values $\tau_p^{\cal A}=6$ and $\tau_p^{\cal A}-\tau_f^{\cal A}=2.75$ have been  found. In physical units,
setting the effective temperature at the end of the quench to $T_{eff}=500$ MeV, the delays correspond to $0.42$ fm/c in model ${\cal B}$, and to $1.03$ fm/c in model ${\cal A}(2)$, which are comparable to the values
inferred from phenomenological analyses of heavy ion collisions.

The metric functions $A(r,\tau)$, $\Sigma(r,\tau)$ and $B(r,\tau)$ appearing in \eqref{metric5D} and computed in \cite{Bellantuono:2015hxa} will be used in the analysis of various nonlocal probes.
Also in these cases, to study thermalization using different observables it is necessary to compare the results with those obtained in the hydrodynamic setup.
The 5$d$ metric reproducing, through the holographic renormalization procedure, the results in (\ref{hydroeps}-\ref{hydropperp}) must be known.
In the case of 5$d$ Fefferman-Graham coordinates, the metric was derived in \cite{Janik:2005zt,Janik:2006ft,Heller:2007qt,Heller:2012je}; in the case of Eddington-Finkelstein coordinates this was done in \cite{vanderSchee:2012qj}.
To have a link with the results for the stress-energy tensor components, the 5$d$ metric dual to viscous hydrodynamics can be written as

\be
ds^2=-A^H(r,\tau) d\tau^2+ [\Sigma^H(r,\tau)]^2 e^{B^H(r,\tau)} dx_\perp^2+ [\Sigma^H(r,\tau)]^2 e^{-2B^H(r,\tau)}dy^2 + 2 dr d\tau \,\,\, , \label{metric5Dhydro}
\ee
with the metric functions expressed in terms of the energy density and pressures:
\bea
A^H(r,\tau) &=& r^2 \left(1 - \frac{4}{3 r^4} \epsilon(\tau) \right) \nn \\
\Sigma^H(r,\tau) &=&r \left(\tau+\frac{1}{r}\right)^{1/3} \label{AdSBH} \\
B^H(r,\tau) &=&\frac{1}{3 r^4} \left(p_\perp (\tau)-p_\parallel (\tau) \right)-\frac{2}{3}\log\left(\tau +\frac{1}{r}\right) \,\,. \nn
\eea
Notice that using \eqref{metric5Dhydro} and \eqref{AdSBH}, the relations \eqref{hydroeps}-\eqref{hydropperp} are reproduced also if a constant is added to the metric function $B^H$: $B^H \to B^H +c$.
In the case of model $\mathcal{B}$ we exploit this freedom and add to $B^H$ the constant $\gamma(\infty)$ in order to take into account the residual effect of the quench that persists in this model at late times.
In the following, the hydrodynamic expressions for the various nonlocal probes are determined using Eqs.~\eqref{metric5Dhydro}, \eqref{AdSBH}, and the expressions \eqref{hydroeps}-\eqref{hydropperp} with   $\Lambda$ determined for each model.
\section{Nonlocal probes of thermalization}\label{sec3}
We now consider a set of nonlocal probes of thermalization of the boundary field theory, the equal-time two-point correlation functions and the Wilson loops of different shapes, in particular circular and rectangular.
Their expressions in the holographic framework are given in the following.

Let us first consider equal-time two-point correlation functions and their geodesic approximation.
According to the AdS/CFT dictionary, a boundary scalar operator $\mathcal{O}(t,\bm{x})$ of conformal dimension $\Delta$ in $d$ dimensions is dual to a bulk field $\phi(t,\bm{x},r)$ with mass $m$ in $(d+1)$ dimensions, with $\Delta=\frac{1}{2}(d+\sqrt{d^2+4 m^2})$.
When an expression of the bulk action is available and the wave equation for $\phi(t,\bm{x},r)$ is solved, the equal-time two-point function $\langle \mathcal{O} (t,\bm{x}) \mathcal{O} (t,\bm{x}') \rangle$ can be determined (in the strong-coupling regime of the boundary theory) starting from the on-shell supergravity action.
For involved bulk geometries, the two-point correlation functions can be computed in the geometric optic limit, in terms of the length $\mathcal{L}$ of the space-like geodesics connecting the two points on the boundary \cite{Balasubramanian:1999zv,Louko:2000tp}:
\begin{equation}\label{geo_approximation}
\langle \mathcal{O} (t,\bm{x}) \mathcal{O} (t,\bm{x}') \rangle \simeq \sum_{\mathrm{geodesics}} e^{-\Delta\, \mathcal{L}} \,\,\, .
\end{equation}
The approximation is effective for boundary theory operators with large conformal dimension, $\Delta\gg 1$.
$\mathcal{L}$ is obtained by extremizing the length of the curves connecting the two points, written generically as
\begin{equation}\label{geo_length}
\mathcal{L} = \int_{P}^Q d\lambda\sqrt{\pm g_{MN}\dot{x}^{M} \dot{x}^{N}},
\end{equation}
in terms of the coordinates $x^{M}(\lambda)$ $(M=1,\dots,d+1)$, the parameter $\lambda$, the boundary points ($P$ and $Q$), the metric $g_{MN}$ and the derivative $\dot{x}^{M}\equiv dx^{M}/d\lambda$ (positive and negative signs in the square root for a space-like or time-like curve).
The geodesic, for which $\mathcal{L}$ is extremal, is determined interpreting the integrand in \eqref{geo_length} as a Lagrangian and solving the corresponding Euler-Lagrange equations.

Another nonlocal probe is the expectation value of Wilson loops.
For a closed contour $\mathcal{C}$, the Wilson loop of the boundary theory is defined as
\begin{equation}\label{WL}
W_{\mathcal{C}}[A]=\frac{1}{N_c}Tr\left(P e^{-ig\oint_{\mathcal{C}}dx^{\mu}A_{\mu}^{a}T^{a}}\right).
\end{equation}
In the strong-coupling limit, the expectation value of \eqref{WL} has a holographic expression \cite{Maldacena:1998im}:
\begin{equation}\label{WL_saddle_point}
\langle W_{\mathcal{C}} \rangle \sim e^{-S_{NG}}
\end{equation}
where $S_{NG}$ is the Nambu-Goto action, the area of the string worldsheet bounded by the curve $\mathcal{C}$:
\begin{equation}
S_{NG}=\frac{1}{2\pi\alpha'}\int d^{2}\xi \sqrt{det\left[g_{MN}\partial_{\alpha}X^{M}\partial_{\beta}X^{N}\right]}\,,
\end{equation}
with $\xi^{\alpha}$ $(\alpha,\beta=1,2)$ the worldsheet coordinates, and $X^{M}\left(\xi^{\alpha}\right)$ the embedding of the surface into the target spacetime.

Two-point correlation functions and the vacuum expectation values of Wilson loops of different shapes, in particular circular and rectangular, can be computed in the holographic setup characterized by the Eddington-Finkelstein coordinates $\left(\tau,\bm{x}_{\perp},y,r\right)$ and the metric (\ref{metric5D}).
To exploit the geodesic approximation \eqref{geo_approximation} for a two-point correlation function, we consider the space-like paths connecting the boundary points $P=\left(t_{0},-\ell/2,x_{2},y\right)$ and $Q=\left(t_{0},\ell/2,x_{2},y\right)$, and extending in the bulk at fixed $(x_{2},y)$.
The coordinate $x_{1}\equiv x$ varies along each curve, the profile of which is described by $\tau(x)$ and $r(x)$. In the middle point $x=0$ the values of $\tau$ and $r$ are 
\begin{equation}\label{midpoint_conditions0}
\tau(0)=\tau_* , \qquad r(0)=r_* \,\,\, .
\end{equation}
Moreover, we require
\begin{equation}\label{midpoint_conditions}
\tau'(0)=r'(0)=0 \,\,\, ,
\end{equation}
with the prime indicating a derivative with respect to $x$.
The conditions
\begin{equation}\label{boundary_conditions}
\tau(-\ell/2)=\tau(\ell/2)=t_{0} , \qquad r(-\ell/2)=r(\ell/2)=r_{0}
\end{equation}
are fulfilled at the boundary.
Eq.~\eqref{midpoint_conditions} is due to the $x\leftrightarrow -x$ symmetry along the $\tau$ and $r$ axes, while Eqs.~\eqref{boundary_conditions} involve the cutoff $r_{0}$ in the bulk coordinate, which is needed for the numerical computation as discussed below.

The length of the curves is given by
\begin{equation}\label{our_geo_length}
\mathcal{L} =\int_{\lambda_1}^{\lambda_2} d\lambda \left(-A(r,\tau)\dot{\tau}(\lambda)^2+2\dot{\tau}(\lambda)\dot{r}(\lambda)+\tilde{\Sigma}(r,\tau)\dot{x}(\lambda)^2\right)^{1/2},
\end{equation}
with $\tilde{\Sigma}(r,\tau)\equiv\Sigma(r,\tau)^2e^{B(r,\tau)}$ and the dot indicating a derivative with respect to $\lambda$.
$\lambda_1$ and $\lambda_2$ correspond to $x(\lambda_1)=-\ell/2$ and $x(\lambda_2)=\ell/2$.
This expression, obtained parametrizing the curves in terms of $\lambda$, is analogous to \eqref{geo_length} and allows one to interpret
\be
L(\dot{x},\tau,\dot{\tau},r,\dot{r})=\left(-A(r,\tau)\dot{\tau}(\lambda)^2+2\dot{\tau}(\lambda)\dot{r}(\lambda)+\tilde{\Sigma}(r,\tau)\dot{x}(\lambda)^2\right)^{1/2}
\ee
as a Lagrangian and $x(\lambda)$ a cyclic variable with conjugate momentum $p_{x}$ conserved along the curve.
The conservation equation, using \eqref{midpoint_conditions0} and \eqref{midpoint_conditions}, can be expressed in terms of the coordinate $x$:
\begin{equation}\label{conservation}
\frac{\tilde{\Sigma}(r,\tau)}{\left(-A(r,\tau)\tau'(x)^2+2\tau'(x)r'(x)+\tilde{\Sigma}(r,\tau)\right)^{1/2}}=\tilde{\Sigma}\left(r_{*},\tau_{*}\right)^{1/2} \,\, .
\end{equation}
The geodesics equations
\begin{eqnarray}\label{eq-geo1}
& & A(r,\tau) \tau''(x) - r''(x) + \left[ - A(r,\tau) \frac{\partial_{\tau} \tilde{\Sigma}(r,\tau)}{\tilde{\Sigma}(r,\tau)} + \frac{1}{2} \partial_{\tau} A(r,\tau) \right] \tau'(x)^2 + \frac{\partial_{r} \tilde{\Sigma}(r,\tau)}{\tilde{\Sigma}(r,\tau)} r'(x)^2 \nonumber \\ & & + \left[ \frac{\partial_{\tau} \tilde{\Sigma}(r,\tau)}{\tilde{\Sigma}(r,\tau)} - A(r,\tau) \frac{\partial_{r} \tilde{\Sigma}(r,\tau)}{\tilde{\Sigma}(r,\tau)} + \partial_r A(r,\tau) \right] r'(x) \tau'(x) + \frac{1}{2} \partial_{\tau} \tilde{\Sigma}(r,\tau) = 0 \nonumber \\
\end{eqnarray}
and
\begin{equation}\label{eq-geo2}
\tau''(x) + \left[ \frac{1}{2} \partial_r A(r,\tau) - \frac{\partial_{\tau} \tilde{\Sigma}(r,\tau)}{\tilde{\Sigma}(r,\tau)} \right] \tau'(x)^2 - \frac{\partial_{r} \tilde{\Sigma}(r,\tau)}{\tilde{\Sigma}(r,\tau)} r'(x) \tau'(x) - \frac{1}{2} \partial_{r} \tilde{\Sigma}(r,\tau) = 0
\end{equation}
are obtained by combining the Euler-Lagrange equations for $\tau$ and $r$ with the conservation equation \eqref{conservation}. The solution $(r(x),\tau(x))$, corresponding to a pair of input values $\left(r_{*},\tau_{*}\right)$, allows one to determine the geodesic length
\begin{equation}\label{geo_length_on_shell}
\mathcal{L}=\int_{-\ell/2}^{\ell/2} dx\frac{\tilde{\Sigma}(r,\tau)}{\sqrt{\tilde{\Sigma}(r_{*},\tau_{*})}} \,\,\, ,
\end{equation}
with the separation $\ell$ deduced from \eqref{boundary_conditions}.
This expression  requires a regularization, which we implement
subtracting from the length \eqref{geo_length_on_shell} the same quantity computed in pure $AdS_5$;   the subtraction is carried out in the range of $r$ in which the numerical solution of the bulk geometry has been  determined, i.e. for $r$ up to a UV scale  $r_0$. For model $\cal B$ the asymptotic constant value  $\gamma(t_0\to\infty)$ of the quench profile is taken into account in the subtraction.
The determination of the distance $\ell$ is provided by the relation $r\left(\ell/2 \right)=r_0$.

\begin{figure}
\begin{center}
\includegraphics[width = .45\textwidth]{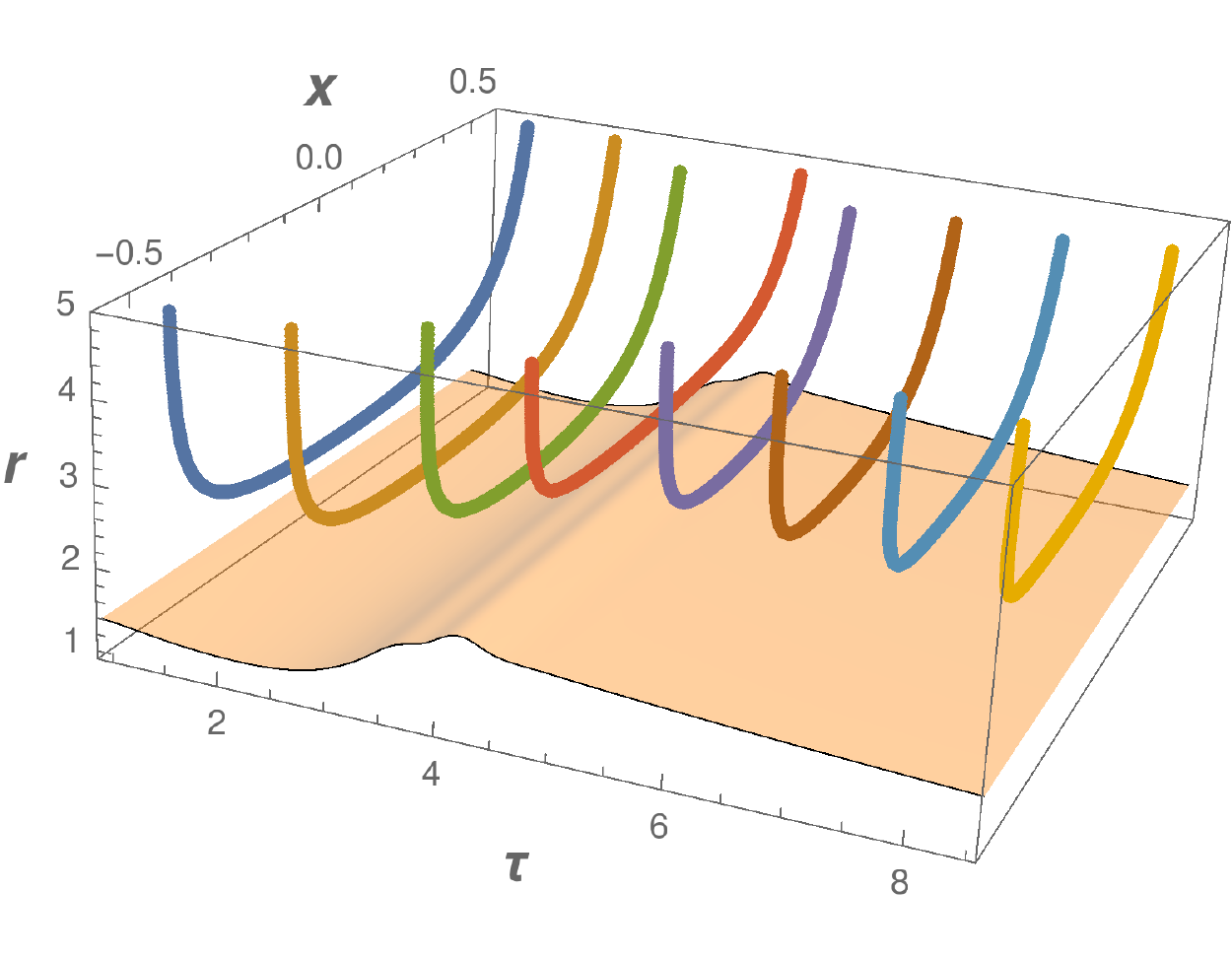}
\caption{\baselineskip 10pt Geodesics obtained in the case of the quench model ${\mathcal B}$ for various $(r_*,\tau_*)$. The shaded area represents the event horizon.}\label{fig:rgeodes}
\end{center}
\end{figure}

Typical resulting geodesics, obtained from the computed metric functions in (\ref{metric5D}), are depicted in Fig.~\ref{fig:rgeodes}.
The thermalization of the boundary theory is studied by computing their lengths as time proceeds.
The hydrodynamic expression of the geodesic lengths $\mathcal{L}_{H}$ are determined in the geometry \eqref{metric5Dhydro}-\eqref{AdSBH} with the same regularization.
This allows to define an observable  by the difference $\Delta\mathcal{L}=\mathcal{L}-\mathcal{L}_{H}$.

\begin{figure}
\begin{center}
\includegraphics[width = .25\textwidth]{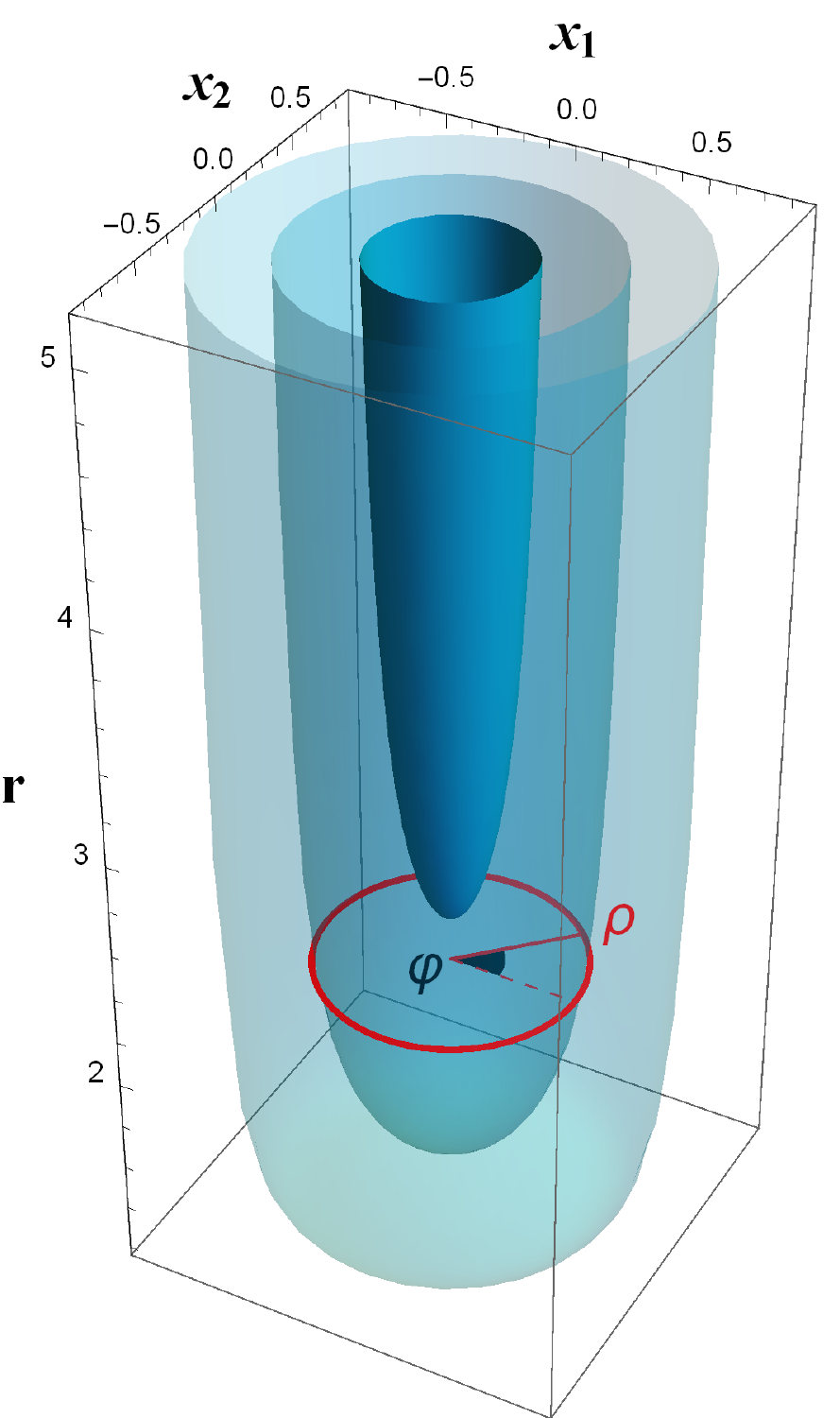} \hspace*{3cm}
\includegraphics[width = .22\textwidth]{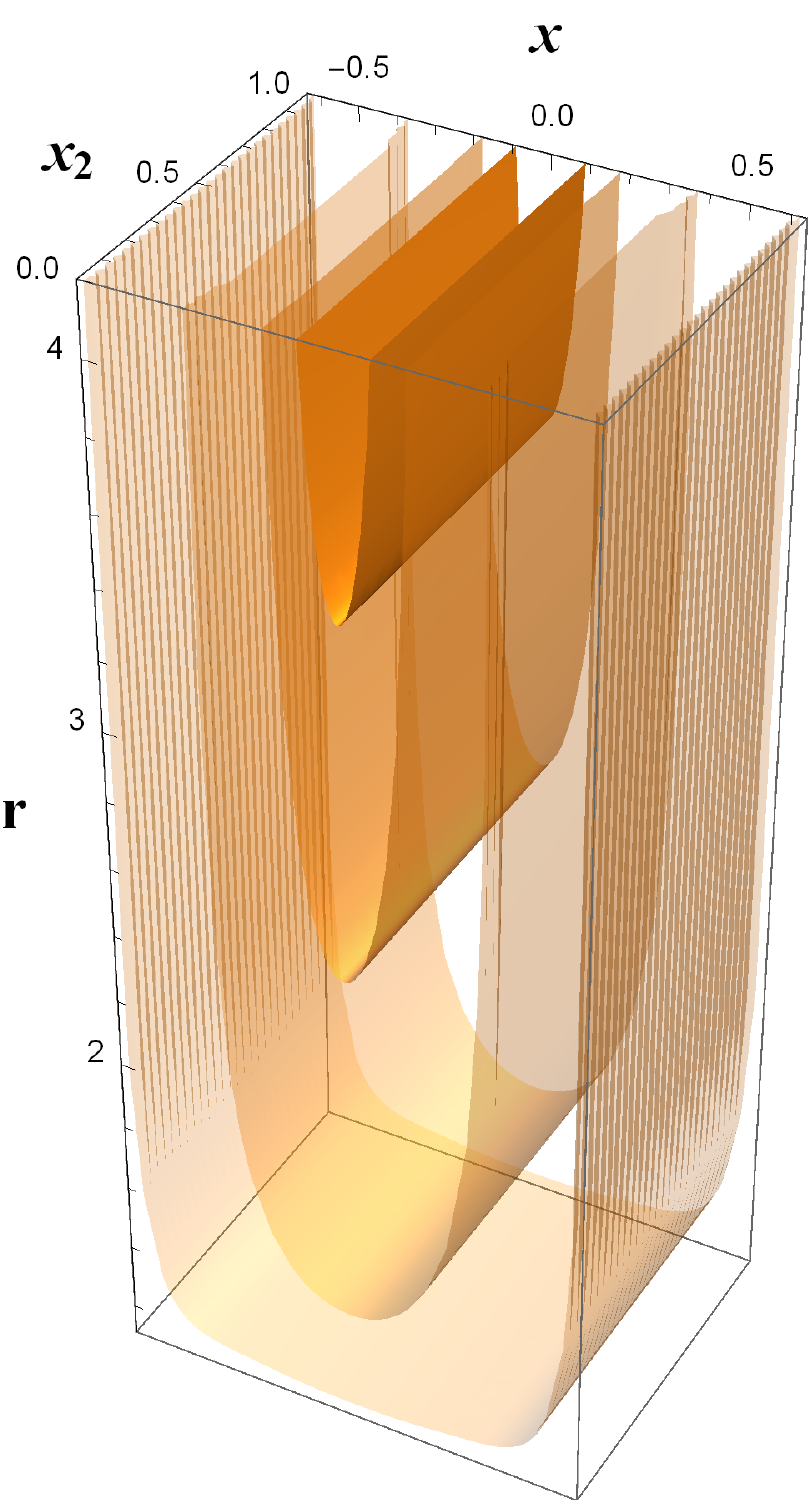}
\caption{\baselineskip 10pt Extremal surfaces of circular (left) and rectangular (right) Wilson loops, computed for quench model ${\mathcal B}$. The value of $\tau_*$ is set to $\tau_*=3$. }\label{fig:circWL}
\end{center}
\end{figure}

In the case of Wilson loops as nonlocal probes of the boundary theory thermalization, we consider two shapes, circles and strips.
For a Wilson loop along a circumference of radius $R=\ell/2$ on the plane $\bm{x}_{\perp}\equiv\left(x_{1},x_{2}\right)$ at the boundary, the space-like worldsheet of minimal area based on the circular path and extending in the bulk at fixed $y$ must be computed.
Such a surface has an azimuthal symmetry and a tip at $\left(\tau,\bm{x}_{\perp},r\right)=\left(\tau_{*},\bm{0},r_{*}\right)$ with $\left(r_{*},\tau_{*}\right)$ input values in the calculation.
The transverse section at fixed $\tau$ and $r$ is a circumference.  For each section the worldsheet can be parametrized in polar coordinates $\xi^{\alpha}=(\rho,\varphi)$, so that
\begin{equation}
\tau=\tau(\rho), \quad x_{1}=\rho\cos\varphi, \quad x_{2}=\rho\sin\varphi, \quad r=r(\rho) \quad , \,\,\, \text{ $y$ fixed.}
\end{equation}
The area of the worldsheet is obtained from the Nambu-Goto action
\begin{equation}\label{WLcircular_Nambu-Goto}
\mathcal{A}_C=\frac{1}{\alpha'}\int_{0}^{\ell/2}d\rho\,\rho\left(\tilde{\Sigma}(r,\tau)\left[-A(r,\tau)\tau'(\rho)^{2}+\tilde{\Sigma}(r,\tau)+2\tau'(\rho)r'(\rho)\right]\right)^{1/2}\,,
\end{equation}
with the prime in the functions $\tau$ and $r$ denoting a derivative with respect to $\rho$, and the angle $\varphi$ integrated out.
Interpreting the integrand of \eqref{WLcircular_Nambu-Goto} as a Lagrangian, one observes that its explicit $\rho$ dependence implies the absence of a conservation equation.
The resulting Euler-Lagrange equations, although involved, can be worked out in a straightforward way.
The solution $(r(\rho),\tau(\rho))$, together with the conditions \eqref{midpoint_conditions0}-\eqref{midpoint_conditions}, can be used to compute the area of the extremal surface, with the same regularization scheme
adopted for the geodesic lengths.
The corresponding quantity in the hydrodynamic geometry is obtained using the metric functions \eqref{AdSBH}, and the  probe of thermalization is given by the difference $\Delta\mathcal{A}_C=\mathcal{A}_C-\mathcal{A}_{C,H}$.
Examples of extremal surfaces of circular Wilson loops, computed using the bulk geometry \eqref{metric5D} for a particular model of quench, are depicted in Fig.~\ref{fig:circWL}.

A less symmetric Wilson loop is an infinite rectangular strip, regarded as a limit of an ellipsoidal loop with an elongated axis.
On the boundary, we set a rectangular contour parametrized by the coordinates $\left(x_{1},x_{2}\right)$, with $-\ell/2 \leq x_1 \leq \ell/2$ and $0 \leq x_2 \leq q$.
The side length $q$ is taken to infinity, and the strip is assumed to be translationally invariant along the $x_{2}$ axis.
The profile of the string surface extending in the bulk at fixed $y$, having the rectangular path as its basis, is described by the embedding $\left(\tau\left(x_{1}\right),r\left(x_{1}\right)\right)$, with the conditions \eqref{boundary_conditions}.
In the following, we denote $x \equiv x_{1}$.
The area of the worldsheet, in terms of the side length $q$ and a parameter $\lambda$, is given by
\begin{equation}\label{WLrectangular_Nambu-Goto}
{\mathcal A}_{R}=\frac{q}{2\pi\alpha'}\int_{\lambda_1}^{\lambda_2}d\lambda\left(\tilde{\Sigma}(r,\tau)\left[-A(r,\tau)\dot{\tau}(\lambda)^{2}+2\dot{\tau}(\lambda)\dot{r}(\lambda)+\tilde{\Sigma}(r,\tau)\dot{x}(\lambda)^{2}\right] \right)^{1/2}\, ,
\end{equation}
with $x(\lambda_{1,2})=\mp \ell/2$.
Interpreting the integrand in \eqref{WLrectangular_Nambu-Goto} as a Lagrangian, $x$ is a cyclic variable with conjugate momentum conserved on the worldsheet.
The conservation equation, using \eqref{midpoint_conditions0}-\eqref{midpoint_conditions}, reads
\begin{equation}\label{WLrectangular_conservation}
\frac{\tilde{\Sigma}(r,\tau)^{3/2}}{\left(-A(r,\tau)\tau'(x)^{2}+2\tau'(x)r'(x)+\tilde{\Sigma}(r,\tau)\right)^{1/2}}=\tilde{\Sigma}\left(r_{*},\tau_{*}\right) \,\,\,
\end{equation}
in terms of the coordinate $x$.
Solving the equations
\begin{eqnarray}\label{eq-wl1}
& & A(r,\tau) \tau''(x) - r''(x) + \left[ - \frac{3}{2}A(r,\tau) \frac{\partial_{\tau} \tilde{\Sigma}(r,\tau)}{\tilde{\Sigma}(r,\tau)} + \frac{1}{2} \partial_{\tau} A(r,\tau) \right] \tau'(x)^2 + \frac{\partial_{r} \tilde{\Sigma}(r,\tau)}{\tilde{\Sigma}(r,\tau)} r'(x)^2 \nonumber \\ & & + \left[ 2\frac{\partial_{\tau} \tilde{\Sigma}(r,\tau)}{\tilde{\Sigma}(r,\tau)} - A(r,\tau) \frac{\partial_{r} \tilde{\Sigma}(r,\tau)}{\tilde{\Sigma}(r,\tau)} + \partial_{r} A(r,\tau) \right] r'(x) \tau'(x) + \partial_{\tau} \tilde{\Sigma}(r,\tau) = 0
\end{eqnarray}
and
\begin{eqnarray}\label{eq-wl2}
& &\tau''(x) + \left[ \frac{1}{2} \partial_{r} A(r,\tau) - \frac{\partial_{\tau} \tilde{\Sigma}(r,\tau)}{\tilde{\Sigma}(r,\tau)} + \frac{1}{2}A(r,\tau)\frac{\partial_{r} \tilde{\Sigma}(r,\tau)}{\tilde{\Sigma}(r,\tau)} \right] \tau'(x)^2 \nn \\
&&- 2\frac{\partial_{r} \tilde{\Sigma}(r,\tau)}{\tilde{\Sigma}(r,\tau)} r'(x) \tau'(x) - \partial_{r} \tilde{\Sigma}(r,\tau) = 0
\end{eqnarray}
allows one to compute the area
\begin{equation}
\mathcal{A}_R=\frac{q}{2\pi \alpha'}\int_{-\ell/2}^{\ell/2}dx\frac{\tilde{\Sigma}(r,\tau)^{2}}{\tilde{\Sigma}\left(r_{*},\tau_{*}\right)} \,\,\, .
\end{equation}
The mentioned regularization scheme is used also for this observable.
The quantity $\mathcal{A}_{RH}$ is computed in the geometry \eqref{AdSBH}, and the difference $\Delta \mathcal{A}_R=(\mathcal{A}_R-\mathcal{A}_{RH})/q$ at various $\tau_{0}$ and for different $\ell$ defines an observable to study thermalization of the boundary theory.
An example of rectangular Wilson loops computed in the geometry \eqref{metric5D}  is shown in Fig.~\ref{fig:circWL}.

\section{Results and discussions}\label{sec4}
We can now  compute the nonlocal observables in the geometry \eqref{metric5D}, with the metric functions numerically determined in \cite{Bellantuono:2015hxa} for the different quench models.
In the three cases, two-point correlation functions and Wilson loops, we have solved the systems of differential equations \eqref{eq-geo1}-\eqref{eq-geo2} and \eqref{eq-wl1}-\eqref{eq-wl2}, together with the equations for the circular Wilson loop. The range $r \leqslant r_0$ is considered for the radial coordinate, with  $r_0=12$.

\subsection{Quench model $\mathcal{B}$}
\begin{figure}[t!]
\begin{center}
\begin{tabular}{cc}
\includegraphics[width = .45\textwidth]{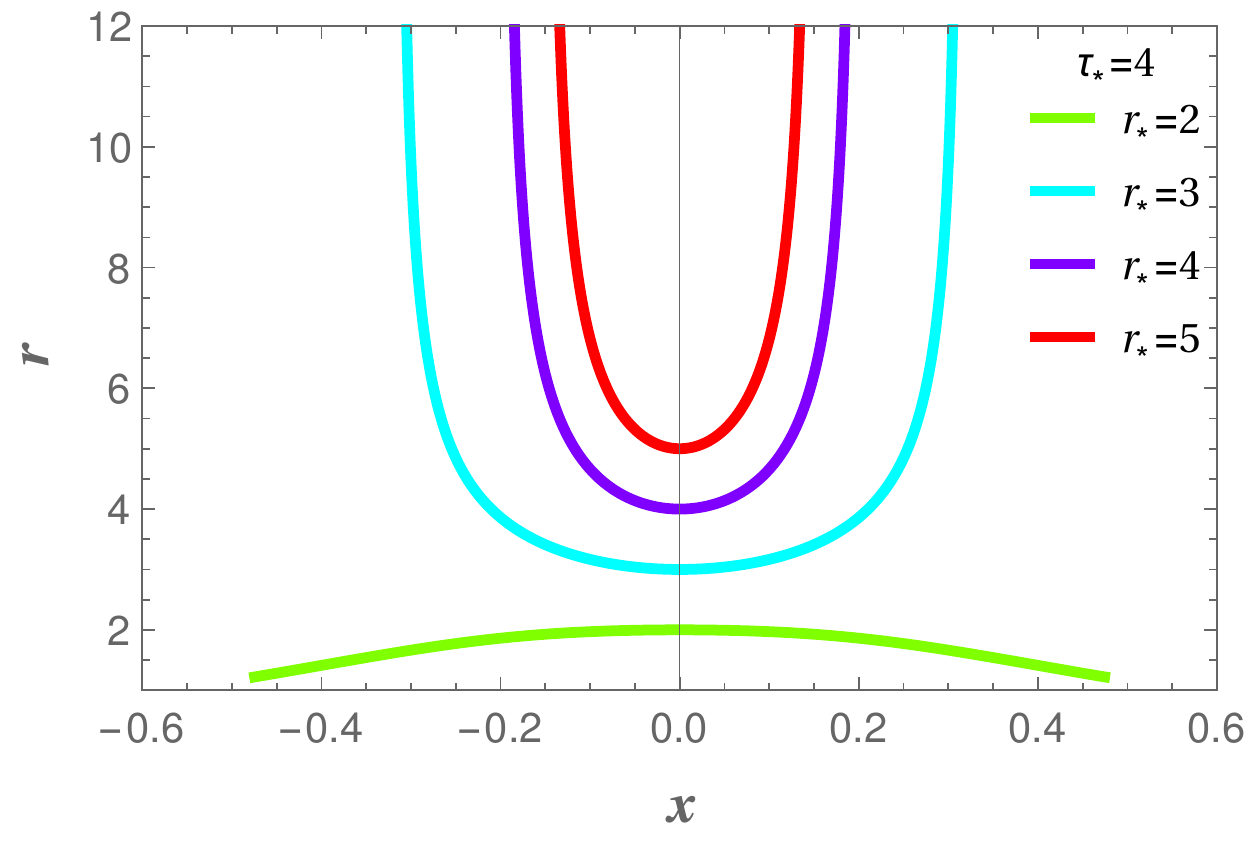} &
\includegraphics[width = .45\textwidth]{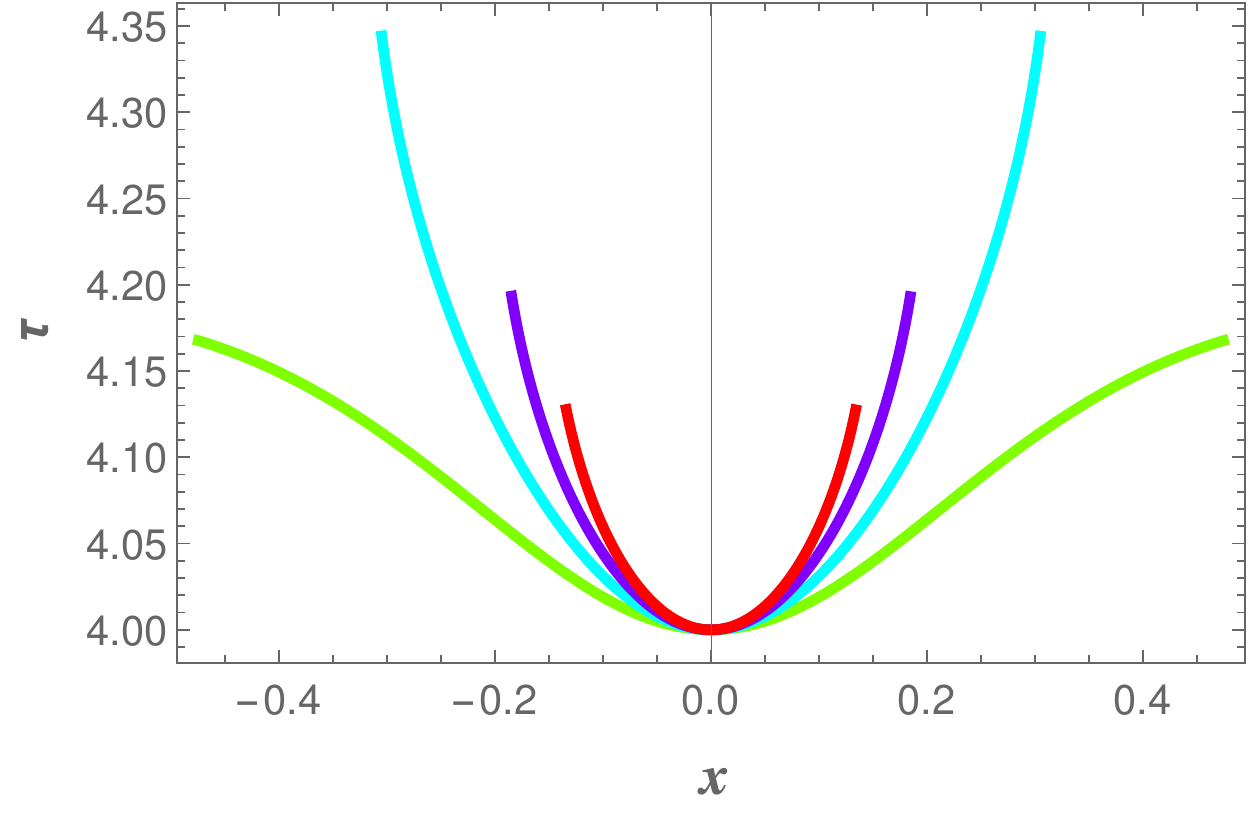} \\
(a)&(b)
\end{tabular}
\caption{\baselineskip 10pt Quench model $\mathcal{B}$. Geodesics $r(x)$ (a) and $\tau(x)$ (b), for $\tau_*=4$ and the values of $r_*$ in the legend.}\label{fig:geort}
\end{center}
\end{figure}

\begin{figure}[t!]
\begin{center}
\includegraphics[width = .45\textwidth]{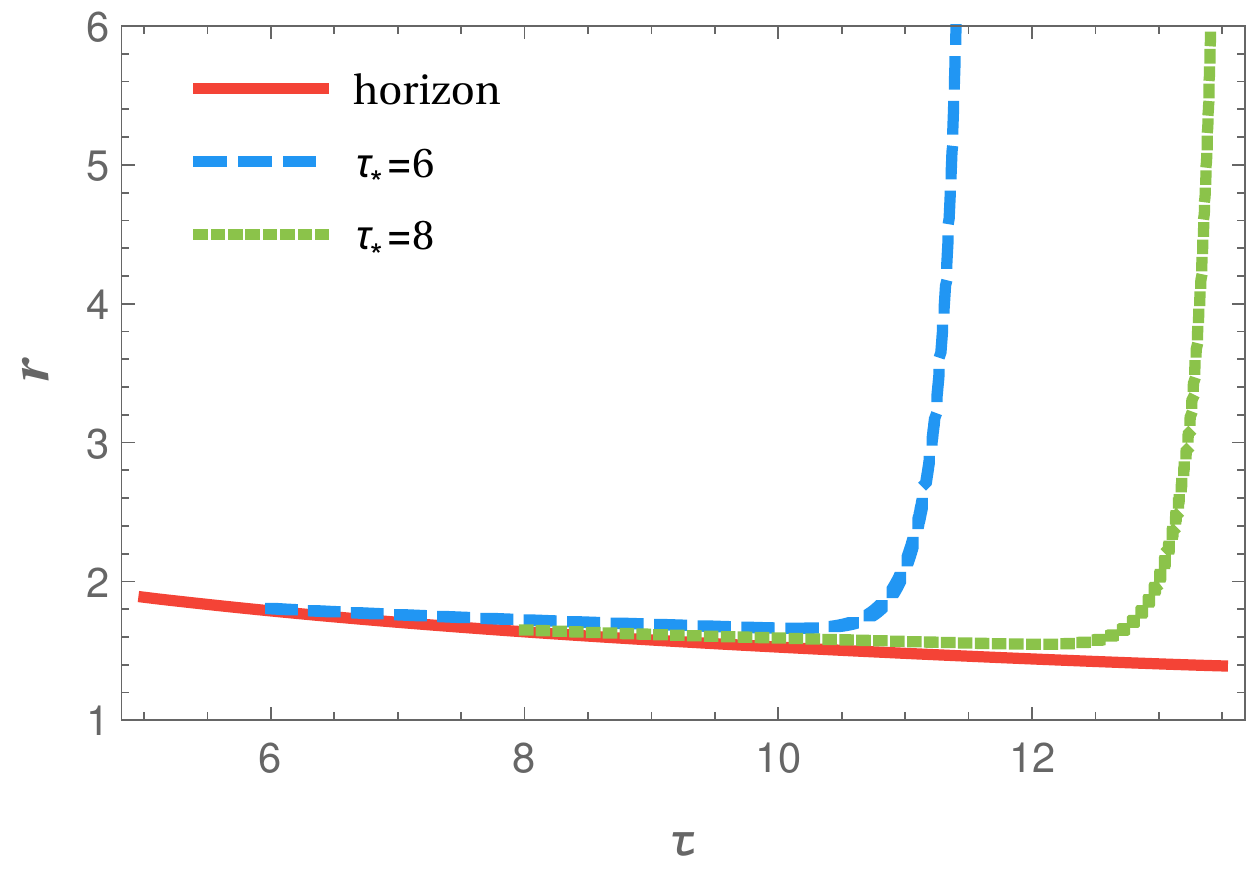} 
\caption{\baselineskip 10pt
Geodesics in the $(\tau,r)$ plane for ($\tau_*=6$, $r_*\sim 1.80$) and ($\tau_*=8$, $r_*\sim 1.65$) in quench model $\mathcal{B}$. Increasing $\tau_*$ (after the pulse in the quench) and for large $\ell$, the radial coordinate closely follows the event horizon.}\label{fig:bh}
\end{center}
\end{figure}

In the case of geodesics and the quench model $\mathcal{B}$, a few solutions $r(x)$ and $\tau(x)$ are shown in Fig.~\ref{fig:geort}.
They are obtained by solving Eqs.~\eqref{eq-geo1}-\eqref{eq-geo2} together with the conditions \eqref{midpoint_conditions0}-\eqref{midpoint_conditions}, with parameters specified in the legendae.
Depending on  $r_*$ and $\tau_*$, two sets of geodesics $r(x)$ are found: those reaching the AdS boundary, the class we are interested in, and those falling into the bulk.
After the quench, at a fixed $\tau_*$, a critical value $r_{*c}$ separates the two classes of solutions, and corresponds to the position of the black brane event horizon.
The solutions at large $\ell$ approach and follow the  horizon, as shown in Fig.~\ref{fig:bh}, and  large boundary separations can be obtained with limits only imposed by the  accuracy of the numerical  algorithm. 
The same $r_{*c}(\tau_*)$ is  found for the  geodesics and  the Wilson loops.

During the quench, when large time gradients are present, we have also found  solutions  starting  from the boundary and crossing the apparent horizon. This phenomenon has been  remarked for nonlocal observables in  rapidly changing  time-dependent setups \cite{Hubeny:2002dg,Hubeny:2012ry,Hartman:2013qma,Liu:2013qca}.

\begin{figure}[t!]
\begin{center}
\begin{tabular}{ll}
 \includegraphics[width = .5\textwidth]{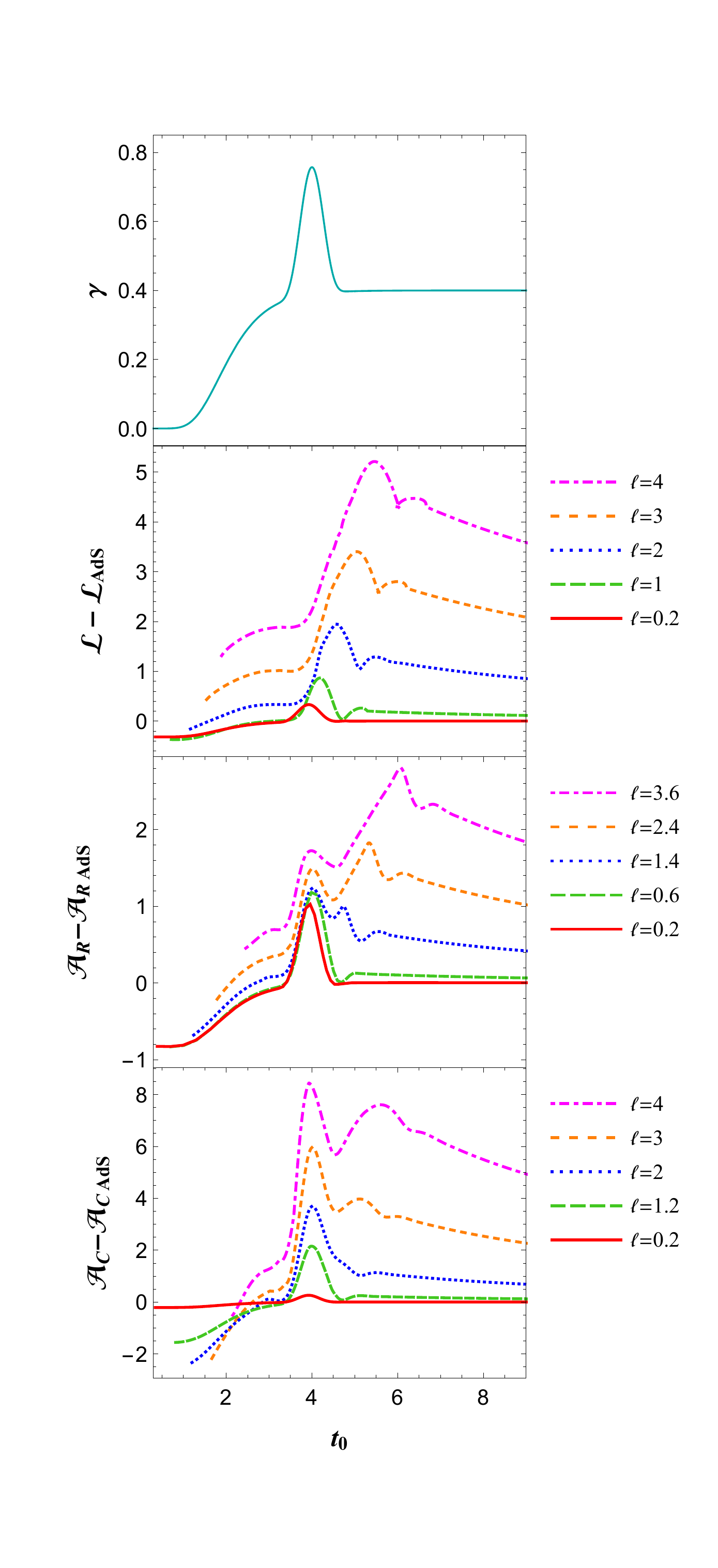} & \hspace*{-1.cm}
 \includegraphics[width = .5\textwidth]{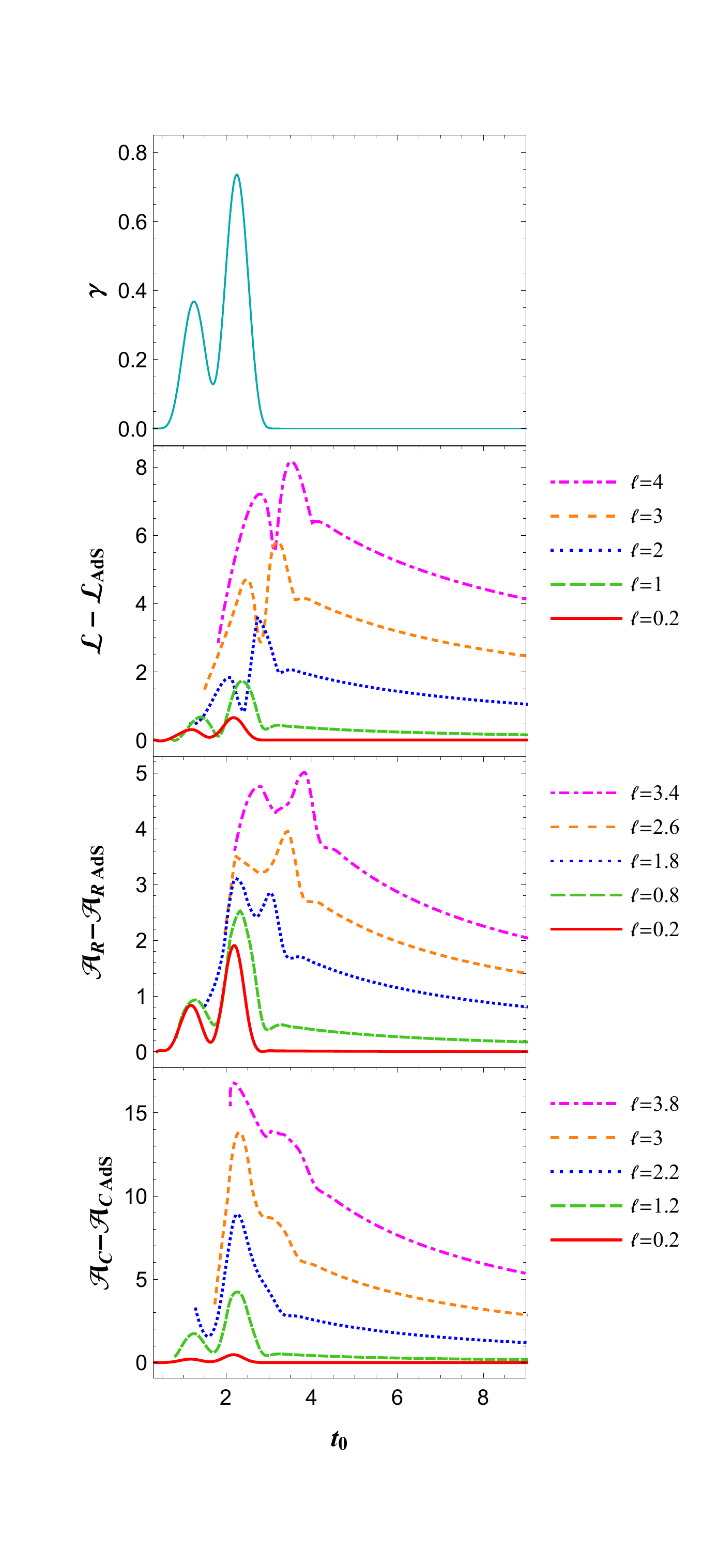}
\end{tabular}
\caption{\baselineskip 10pt Results for quench models $\mathcal{B}$ (left) and $\mathcal{A}(2)$ (right). From top down: profile of the quench $\gamma$, geodesic regularized lengths, regularized areas of the extremal surfaces for rectangular (divided by $q$) and circular Wilson loops versus $t_0$, for the sizes of the probes specified in the legendae. The regularization scheme consists in  subtracting from each observable the corresponding quantity computed in pure $AdS_5$.}\label{fig:lgeo}
\end{center}
\end{figure}

Let us discuss in more detail the results for the regularized geodesic length $\mathcal{L}(t_0,\ell)$, the regularized area of extremal surfaces for the rectangular Wilson loop $\mathcal{A_R}(t_0,\ell)$ (divided by $q$) and for the circular Wilson loop $\mathcal{A_C}(t_0,\ell)$ in model $\mathcal{B}$. They  are shown in Fig.~\ref{fig:lgeo} for several values of the distance $\ell$ between the points in the correlation function, of the side (again denoted by $\ell$) of the rectangular Wilson loop, and of the diameter $\ell$ of the circular Wilson loop ($\alpha'$ is  set to 1).
The curves start at different values of the initial time $t_0$, all corresponding to $\tau_*=0.25$.
 In the plots we limit to $\ell  \simeq 4$, but it is possible to achieve higher values of $\ell$ by increasing the numerical precision. 

\begin{figure}
\centering 
\includegraphics[width=0.5\textwidth]{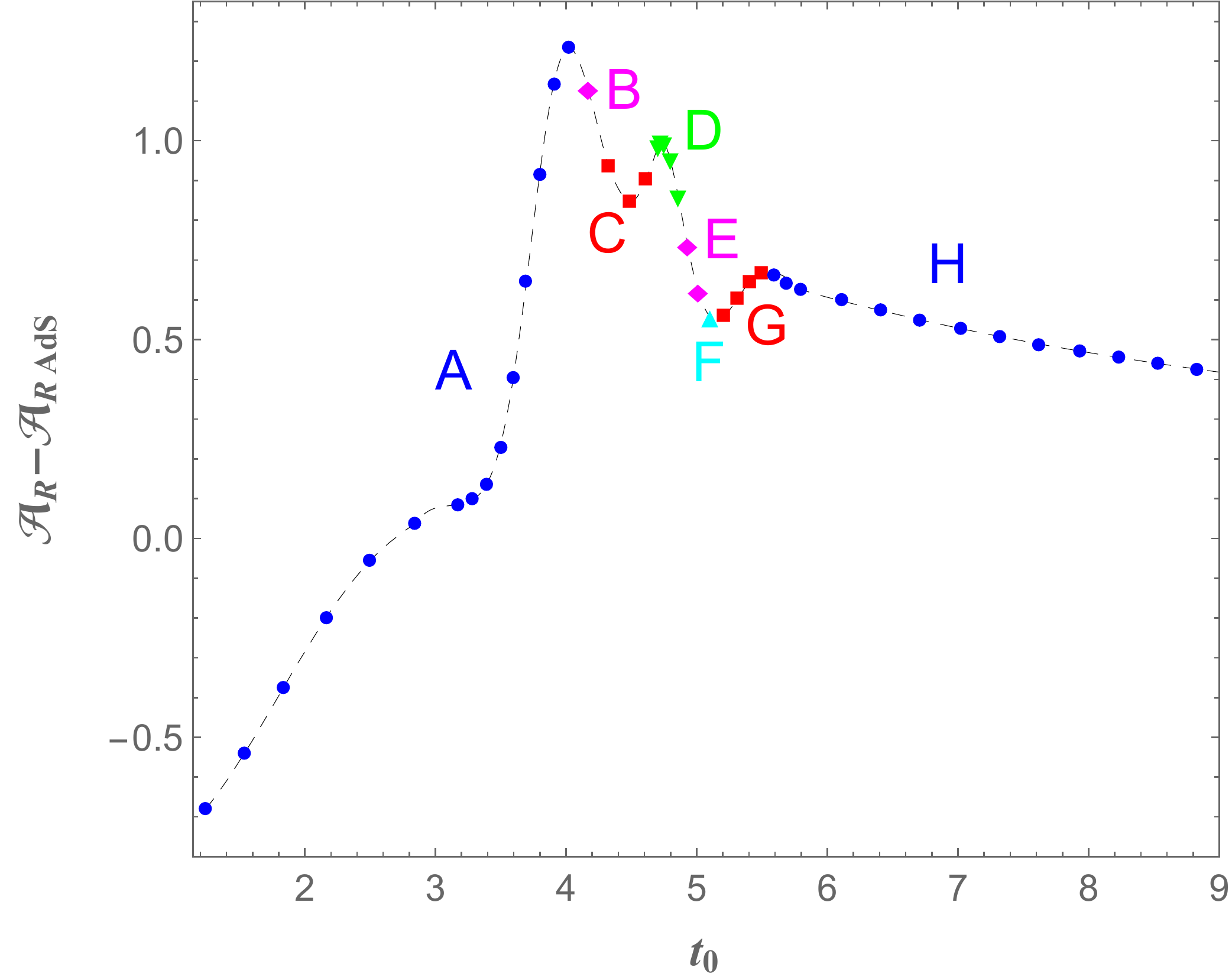}\\
\hspace*{0.85cm} \subfigure[ \, points in A]{\includegraphics[width=0.14\textwidth]{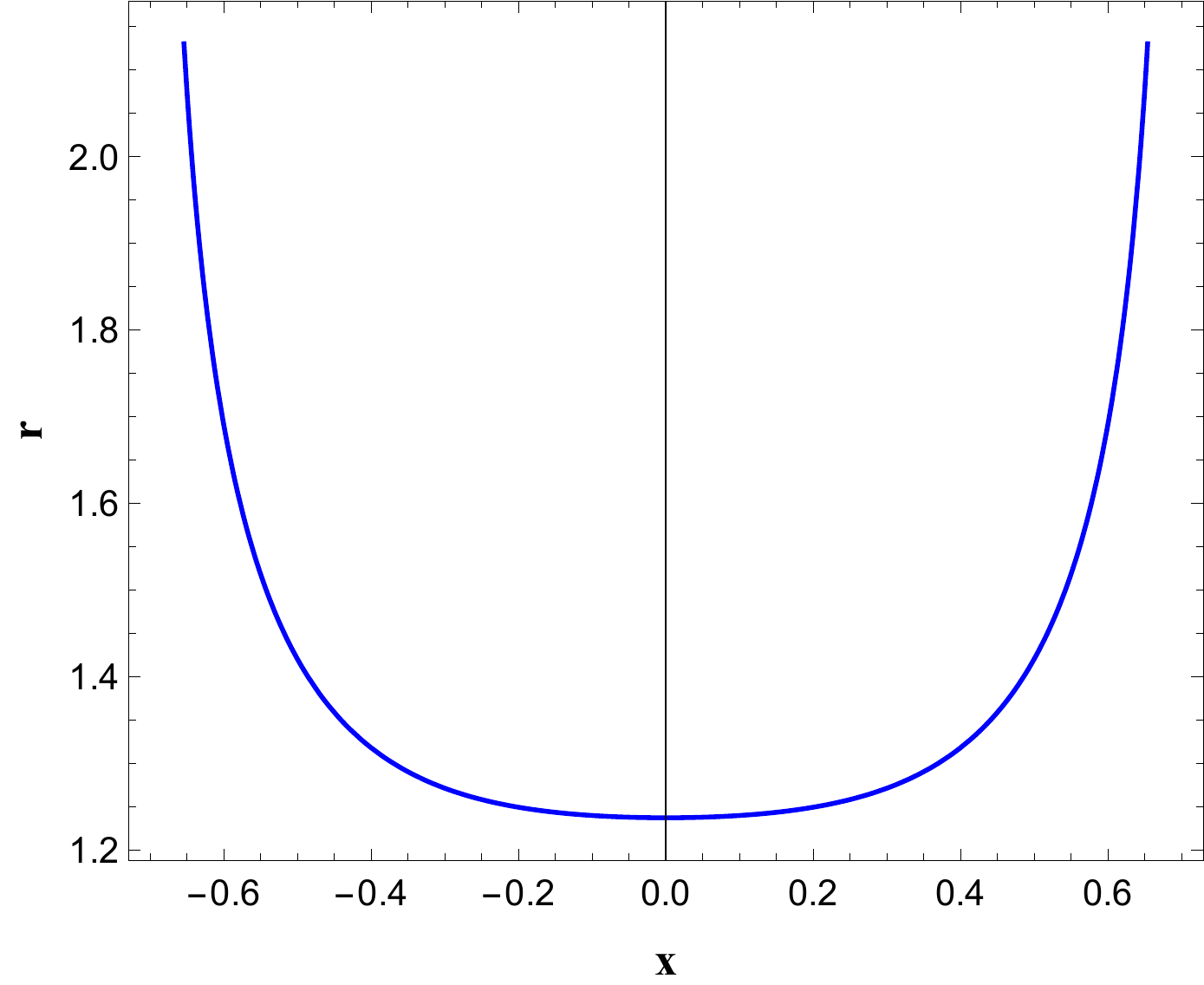}} 
\subfigure[ \, in B]{\includegraphics[width=0.14\textwidth]{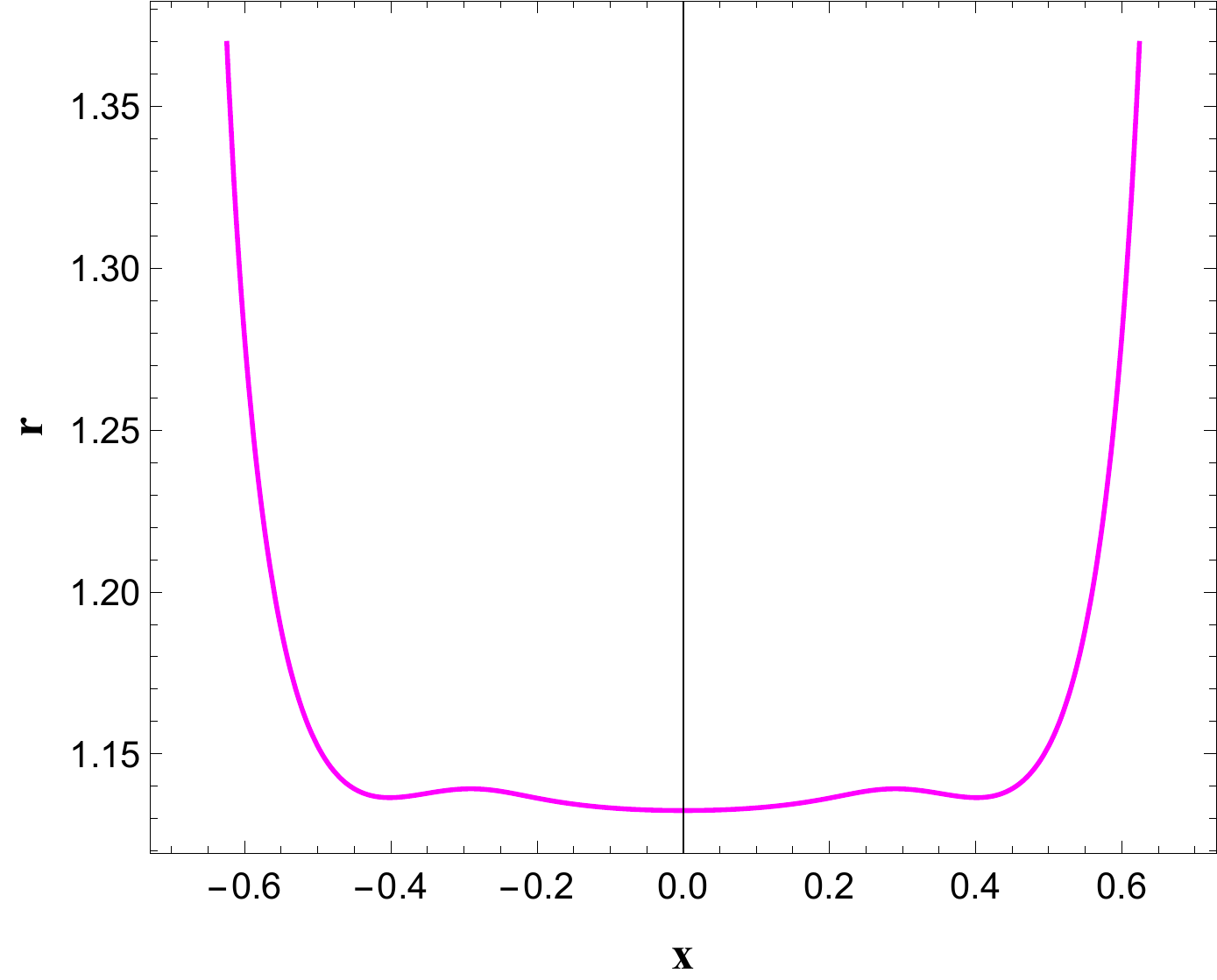}}
\subfigure[ \, in C]{\includegraphics[width=0.14\textwidth]{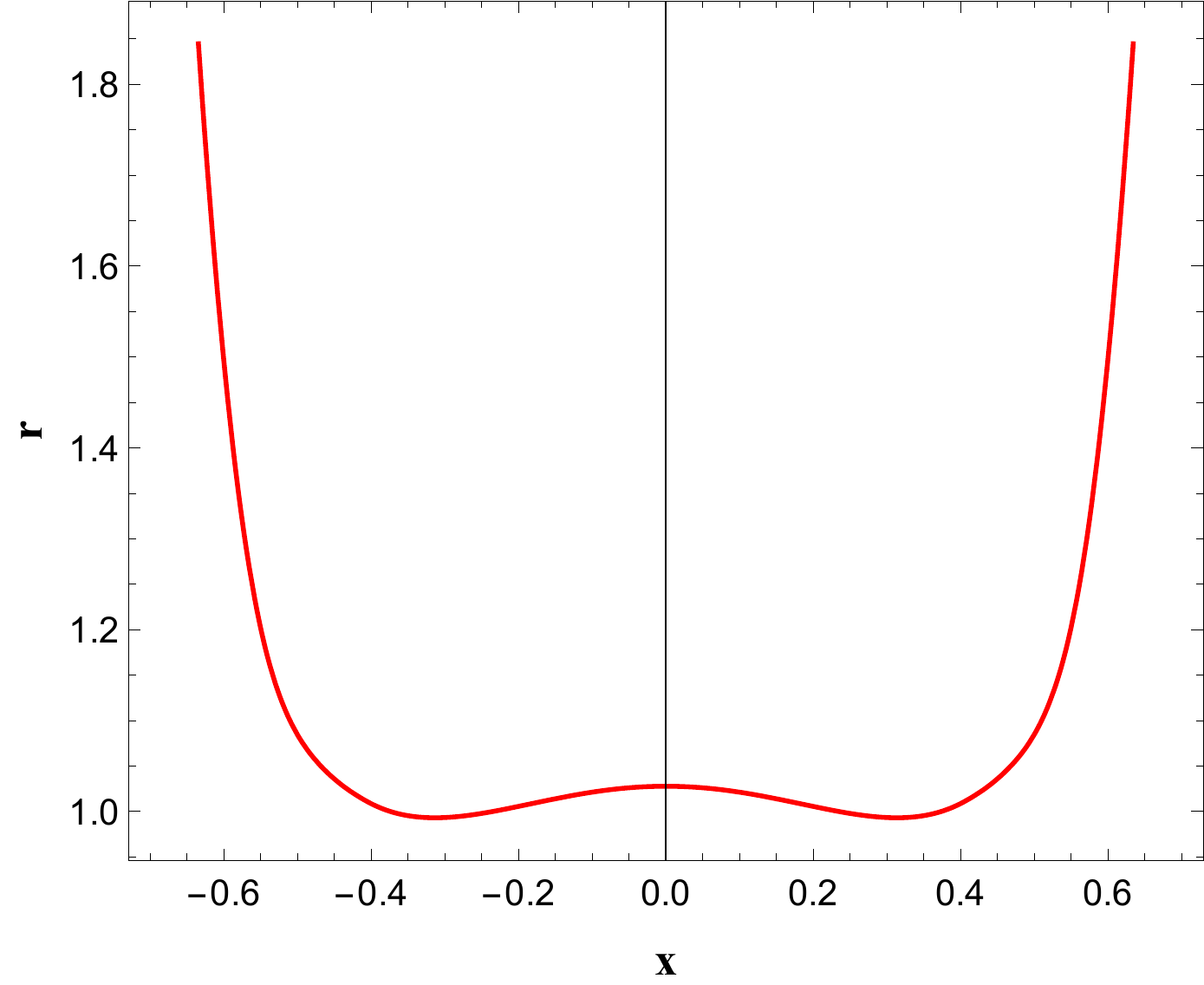}}
\subfigure[ \, in D]{\includegraphics[width=0.14\textwidth]{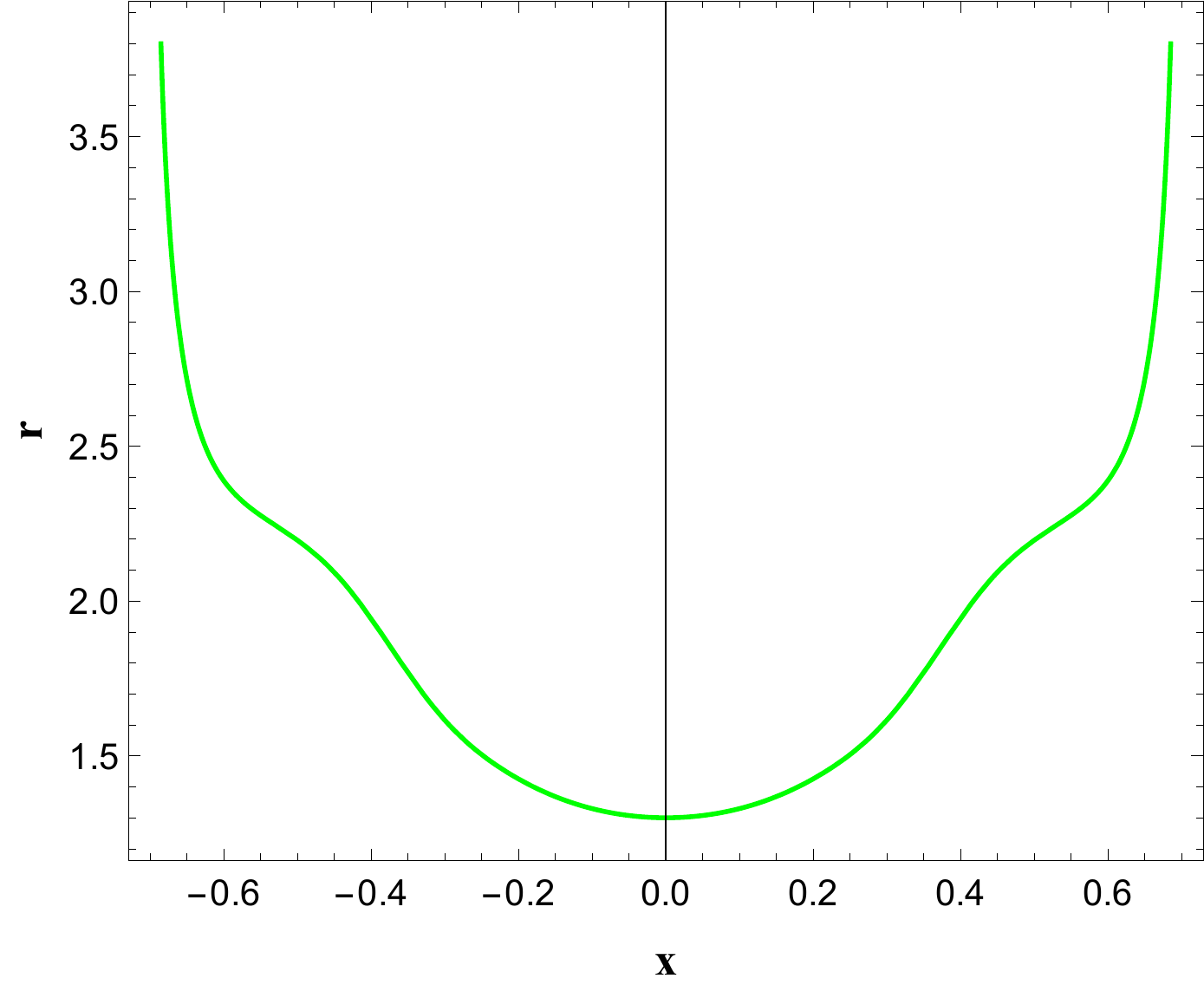}}\\
\hspace*{0.85cm} \subfigure[ \, in E]{\includegraphics[width=0.14\textwidth]{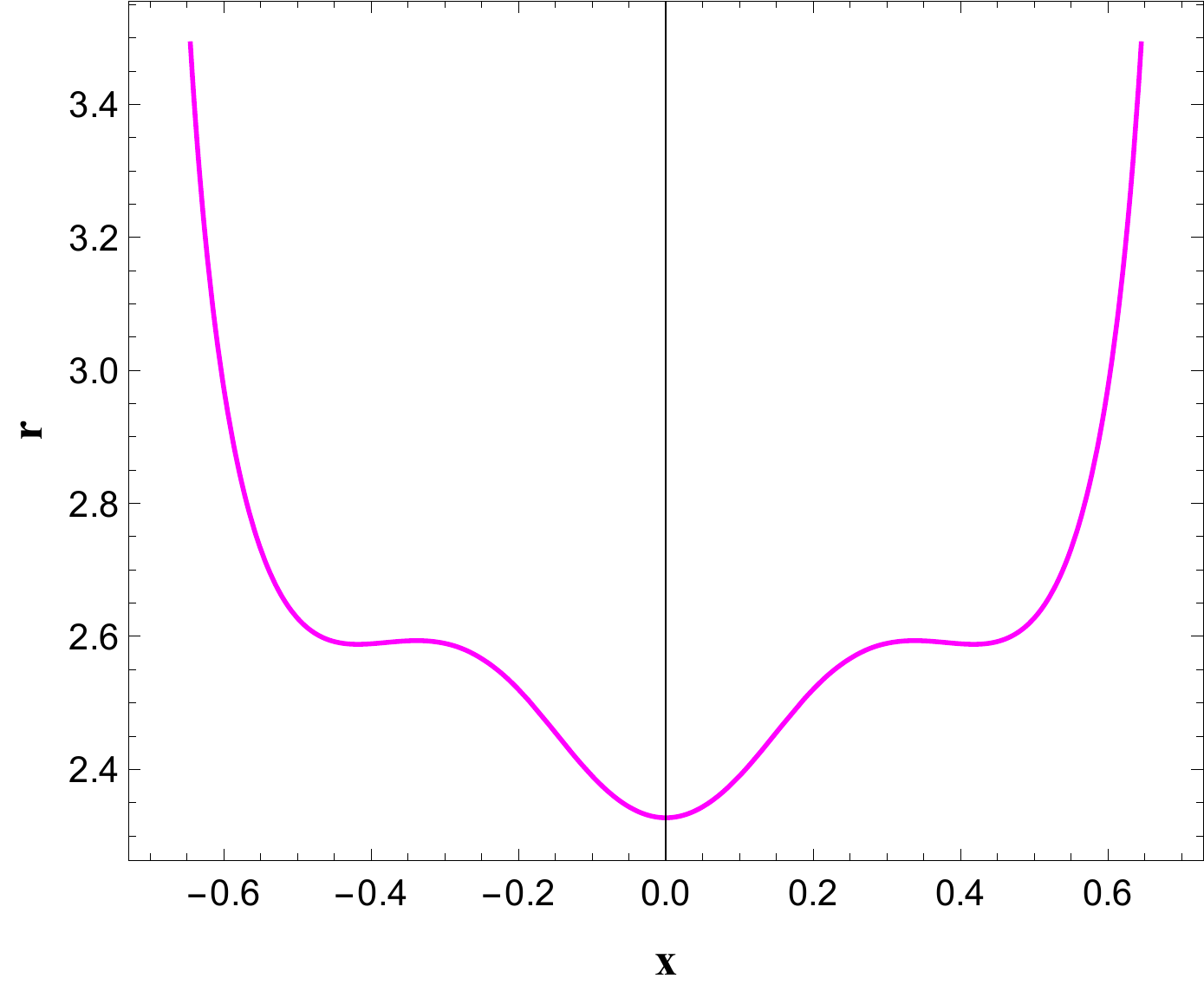}}
\subfigure[ \, in F]{\includegraphics[width=0.14\textwidth]{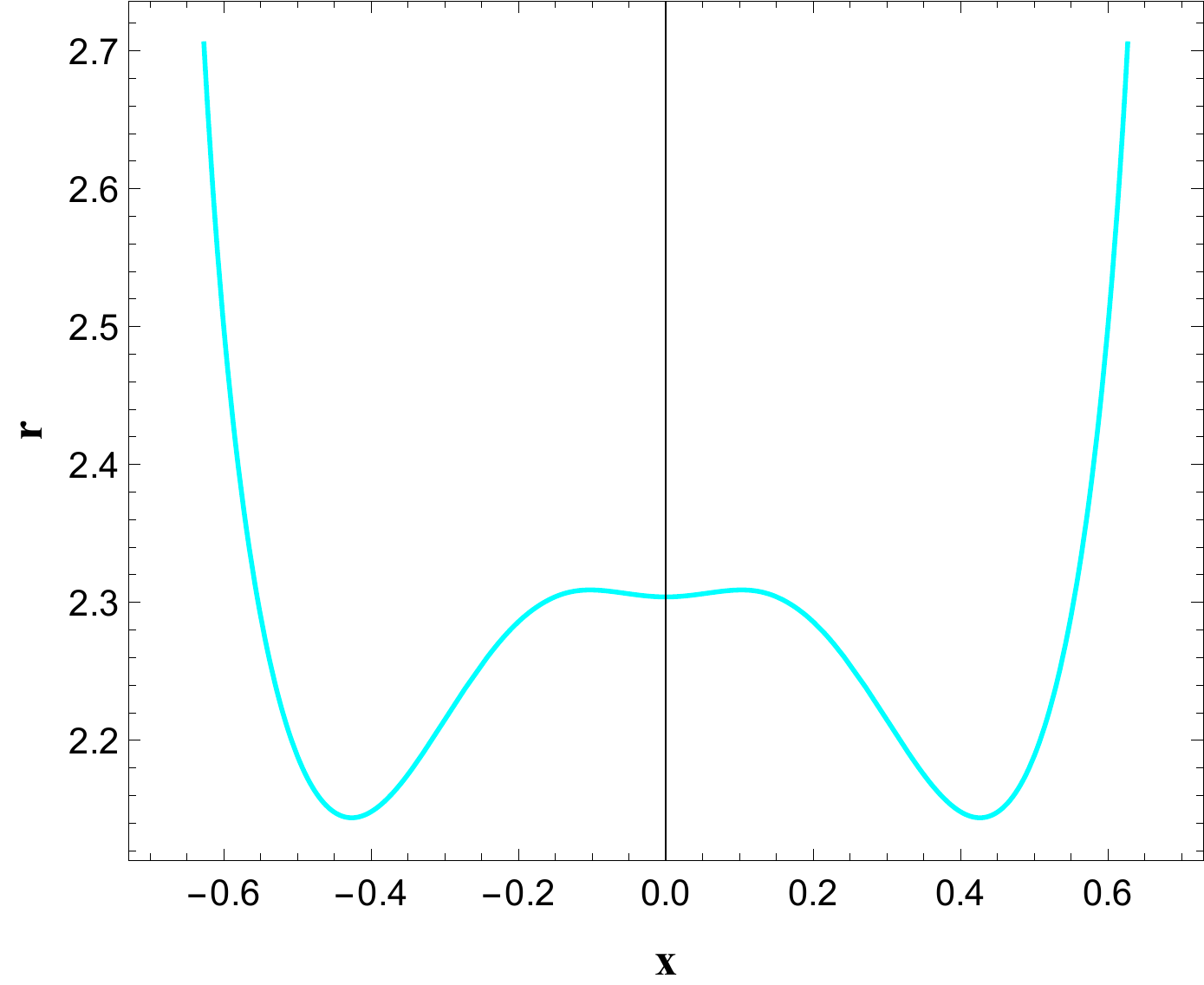}}
\subfigure[ \, in G]{\includegraphics[width=0.14\textwidth]{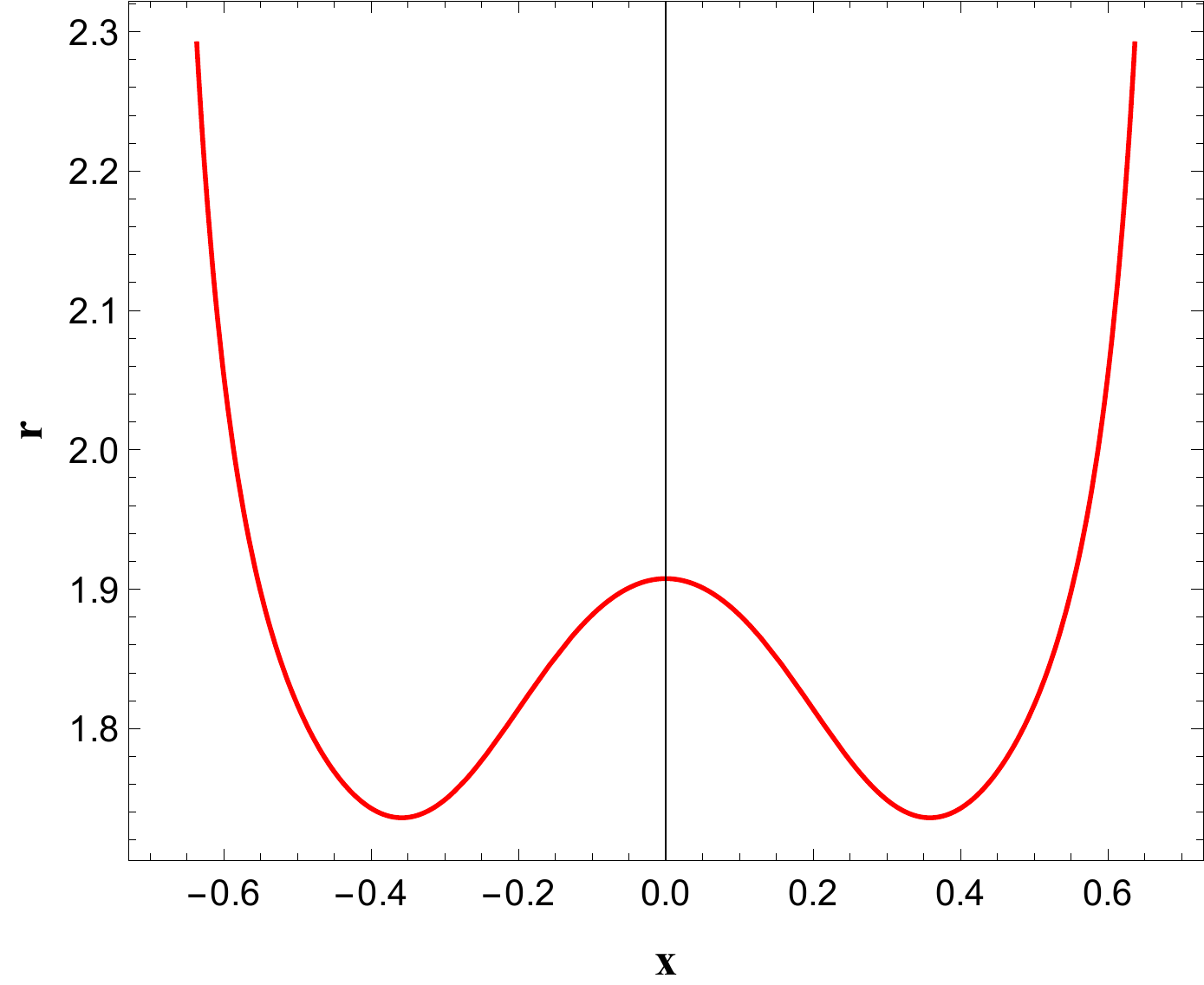}}
\subfigure[ \, in H]{\includegraphics[width=0.14\textwidth]{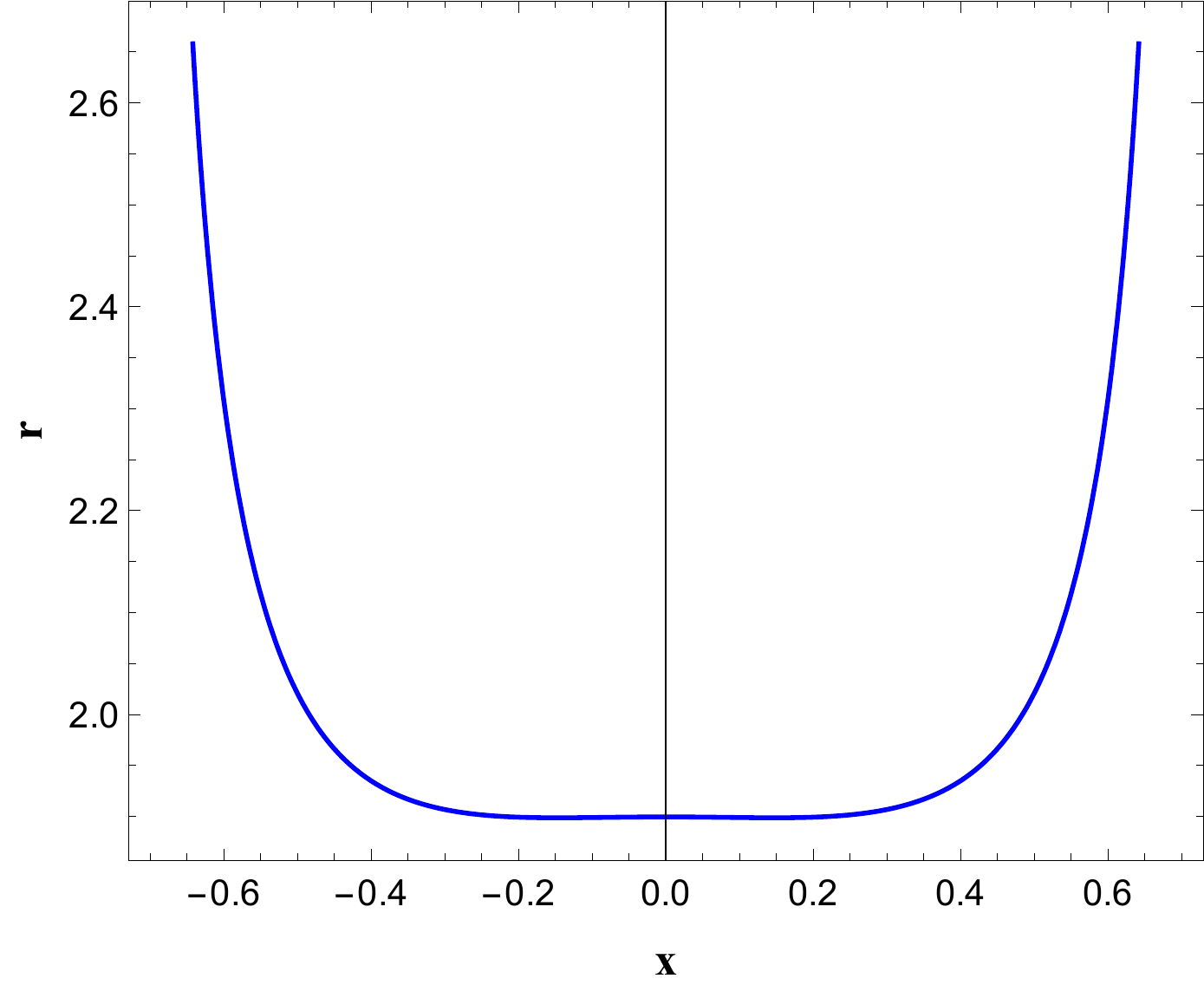}}
\caption{Regularized $\mathcal{A}_{R}$ for model $\mathcal{B}$ and $\ell=1.4$ (top panel). The points with the same color,  in various branches in the plot indicated by a capital letter  from A to H,  correspond to   the  profiles of the solutions $r(x)$  in the corresponding panels from (a) to (h)   in the second and third row.}\label{fig:curves}
\end{figure}

Fig.~\ref{fig:lgeo} shows that the observables follow the quench profile, with a delay that increases for increasing sizes of the probes. 
Additional structures, namely local minima, are found.
They are due to the different profiles of the geodesics and of the extremal surfaces that enter in the bulk.  An example is  reported in Fig.~\ref{fig:curves}, in which we collect the 
profiles of the extremal surfaces for the rectangular Wilson loop as the time proceeds, at a fixed value of $\ell$,   showing that the structures in the regularized area are related to the topologies of the extremal surfaces. Moreover, for such a value of $\ell$ and for values of $t_0$ corresponding to the largest time variations of the geometry (and consequently of the area), extremal surfaces exceeding the event horizon appear, i.e. for some points in branches A (upper part), B, C, D, G.

\begin{figure}[t!]
\begin{center}
\begin{tabular}{ll}
\includegraphics[width = .37\textwidth]{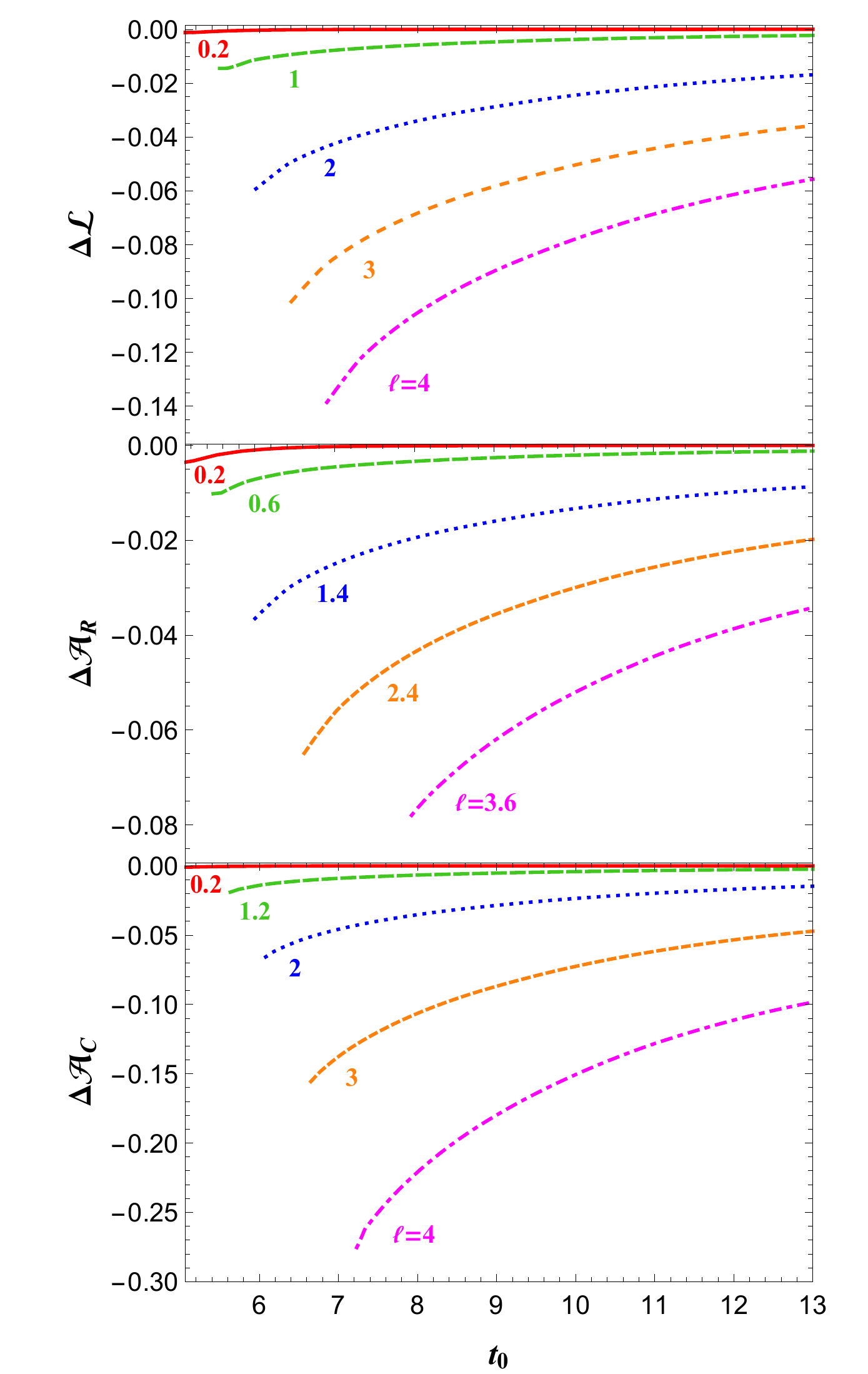} &
\includegraphics[width = .37\textwidth]{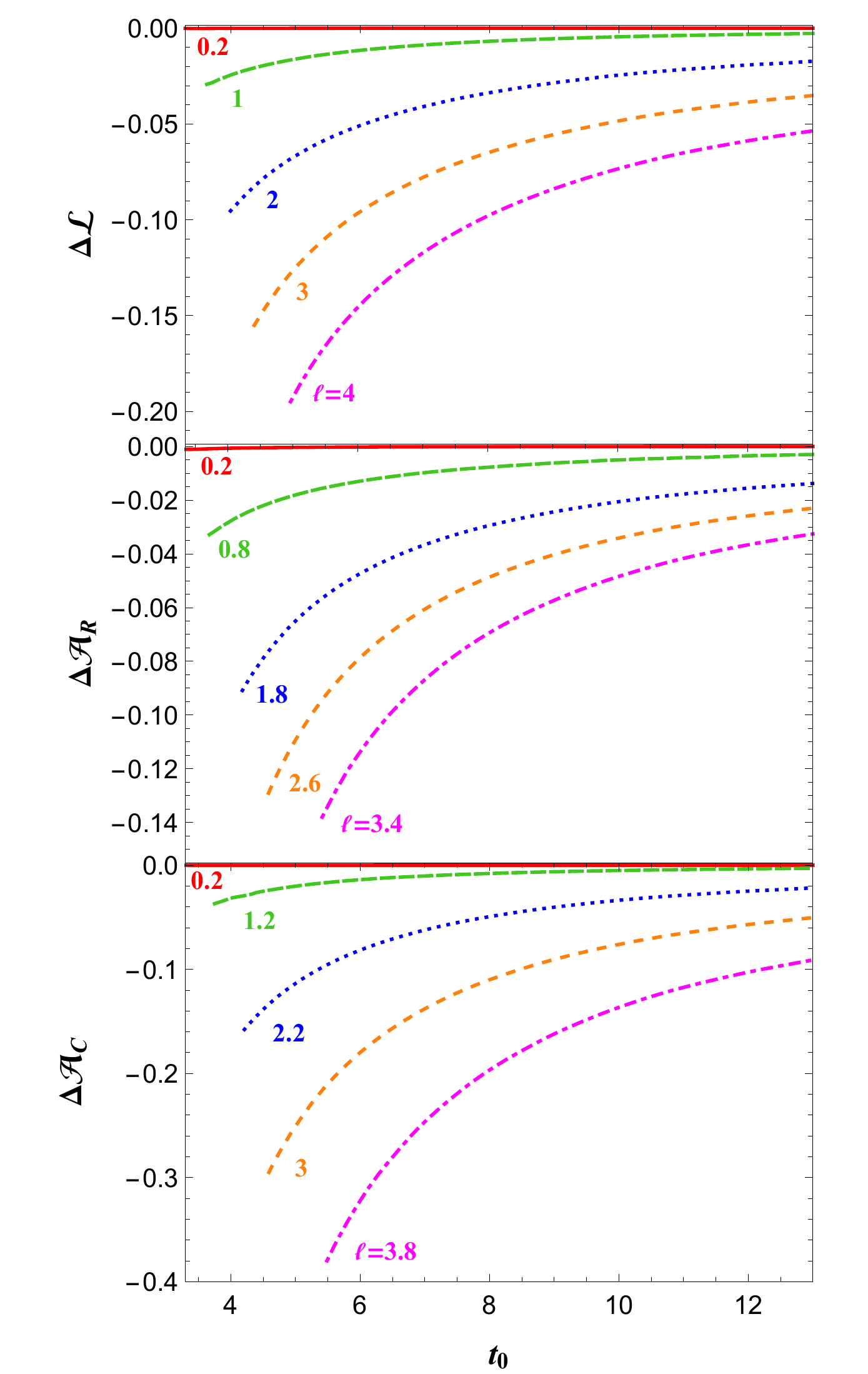}
\end{tabular}
\caption{\baselineskip 10pt Results for quench models $\mathcal{B}$ (left) and $\mathcal{A}(2)$ (right). From top down: Difference between the regularized geodesic length $\Delta\mathcal{L}$ (top), the regularized area (divided by $q$) of the extremal surface for the rectangular Wilson loop $\Delta\mathcal{A}_R$ (middle), and the regularized area of the extremal surface for the circular Wilson loop $\Delta\mathcal{A}_C$ (bottom) in the models with quench and using the hydrodynamic metric. The time $t_0$ starts after the end of the pulse in the quench. }\label{fig:modB}
\end{center}
\end{figure}

Let us focus on the time region that follows the end of the spike in the quench, when the profile $\gamma(\tau)$  is nearly constant.
We are interested in understanding if the nonlocal observables follow the hydrodynamic behavior and, in that case, how fast such a regime is reached after the end of the quench, in comparison with the thermalization time determined through local observables (in particular the pressures).
In Fig.~\ref{fig:modB} we display the differences of the quantities computed in the metric \eqref{metric5D}, and the same quantities computed in the hydrodynamic geometry \eqref{metric5Dhydro}, i.e. the observables $\Delta\mathcal{L}$, $\Delta\mathcal{A}_{R}$ (divided by $q$) and $\Delta\mathcal{A}_{C}$.
The curves in the left panel start at the different values of $t_0$ corresponding to $\tau_*=\tau_f^\mathcal{B}=5$.
This  is due to the fact that, since geodesics are characterised by $\tau_*\leqslant\tau(x)\leqslant t_0$, only the geodesics with $\tau_*\geqslant 5$ are not affected by the quench and can be compared with hydrodynamics.
As shown in Fig.~\ref{fig:modB}, each observable thermalizes at different times where all differences vanish.
The thermalization times are different for different sizes of the probes.
This result, more general than the one found in \cite{Bellantuono:2015hxa}, indicates how nonlocal observables recover the hydrodynamic regime after the end of the quench in comparison with the local observables: $\Delta\mathcal{L}$, $\Delta\mathcal{A}_R$ and $\Delta\mathcal{A}_C$ are smaller for low values of $\ell$, therefore the system is seen to thermalize faster using observables remaining as local as possible.

 Other  remarks are in order.  
An analytic expression for $\mathcal{L}_{Hydro}$ can be obtained in
the  small $\ell$ limit and large $t_0$  \cite{Janik:2005zt,Pedraza:2014moa}. The
leading-order correction with respect to $AdS_5$  is
\begin{equation}
\mathcal{L}_{Hydro}-\mathcal{L}_{AdS}=\frac{\ell^3 \pi^4 (10 + \ell^2
r_0^2)}{120 r_0 \sqrt{4 + \ell^2 r_0^2}} t_0^{-4/3} \Lambda^{8/3} \, + \dots ,
\end{equation}
which has a finite limit for $r_0 \to \infty$  that scales as $\ell^4$ and coincides with the one  in \cite{Pedraza:2014moa}.
A similar expression  holds for the rectangular Wilson loop.

\begin{figure}[b!]
\begin{center}
\includegraphics[width = .45\textwidth]{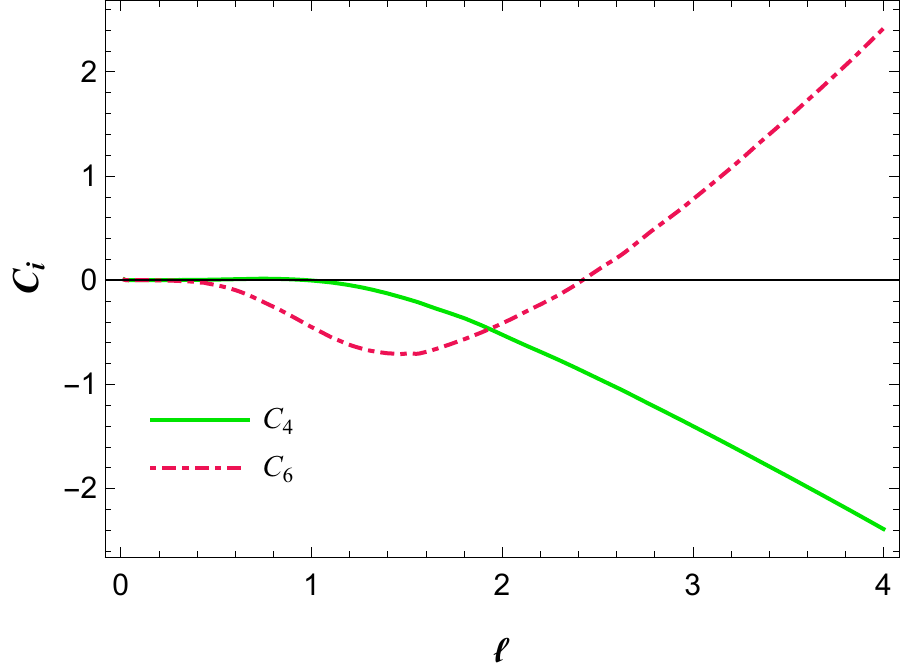}
\caption{\baselineskip 10pt First coefficients of the large $t_0$ expression of the geodesic length  $\dd \Delta {\cal L}=\frac{C_4(\ell)}{(\Lambda t_0)^{4/3}}+\frac{C_6(\ell)}{(\Lambda t_0)^{6/3}}$  for model $\mathcal{B}$. }\label{fig:C4C6}
\end{center}
\end{figure}

As for the difference of the nonlocal observables with respect to hydrodynamics, the  behaviour of the curves in Fig.~\ref{fig:modB} can be  described by the form
$\dd \Delta {\cal L}=\frac{C_4(\ell)}{(\Lambda t_0)^{4/3}}+\frac{C_6(\ell)}{(\Lambda t_0)^{6/3}}$, and similarly for Wilson loops. The first term accounts for a residual $t_0^{-4/3}$
dependence of $\Delta\mathcal{L}$,  which is negligible
for small values of $\ell$  while it increases with  $\ell$ (analogous results hold for  $\mathcal{A}_R$ and $\mathcal{A}_C$). The coefficients $C_4(\ell)$ and $C_6(\ell)$  in Fig.~\ref{fig:C4C6} are very close  for the two models,  showing that after the quench  the nonlocal probes share common features.
They  can be represented in a rational form $C(\ell)=(a \; \ell^b+c)/(d+\ell^e)$ which seems  common to all the observables. For the geodesic length in model $\cal B$ we find
$a=6.58$, $b=-0.03$, $c=-6.59$,  $d=0.07$,  $e=-2.29$ for $C_4(\ell)$,  and 
$a=0.19$, $b=1.70$,  $c=-0.86$,  $d=0.47$, $e=-3.40$ for $C_6(\ell)$.

To provide a quantitative measure of thermalization for the nonlocal probes, we use several criteria to determine the value of the size $\ell$ above which the observables are not thermalized, at a fixed value $t_0=6.74$, corresponding to the restoration of pressure isotropy \cite{Bellantuono:2015hxa}.
\begin{itemize}
 \item $\hat{\ell}_{1}$ is the value of $\ell$ corresponding to the inflection point of the curves $\Delta \mathcal{L}/\ell$, $\Delta \mathcal{A}_R/\ell$ and $\Delta \mathcal{A}_C/\ell^2$ versus $\ell$ at fixed $t_0=6.74$. In the quench model $\mathcal{B}$ the two-point correlation function is thermalized for $\ell \lesssim \hat{\ell}_1=1.0$, the rectangular Wilson loop for $\ell\lesssim \hat{\ell}_1=0.5$ and the circular Wilson loop for $\ell\lesssim \hat{\ell}_1=1.0$. Notice that $\hat{\ell}_{1}$ is almost constant at varying $t_0$, since curves with different $t_0$ have close inflection points, Fig.~\ref{fig:flex}. As $t_0$ increases, the curves have smaller asymptotic slopes which vanish at $t_0\to\infty$.
\item
$\hat{\ell}_2$ is the length corresponding to the inflection point of the derivatives of $\Delta \mathcal{L}$, $\Delta \mathcal{A}_R$ (divided by $q$) and $\Delta \mathcal{A}_C/\ell$. The results $\hat{\ell}_2= 0.7$ (for the two-point correlator), $\hat{\ell}_2= 0.3$ (for the rectangular Wilson loop) and $\hat{\ell}_2= 0.7$ (for the circular Wilson loop) are close to the findings obtained using $\hat{\ell}_1$.
 \begin{figure}[b!]
\begin{center}
\includegraphics[width = .45\textwidth]{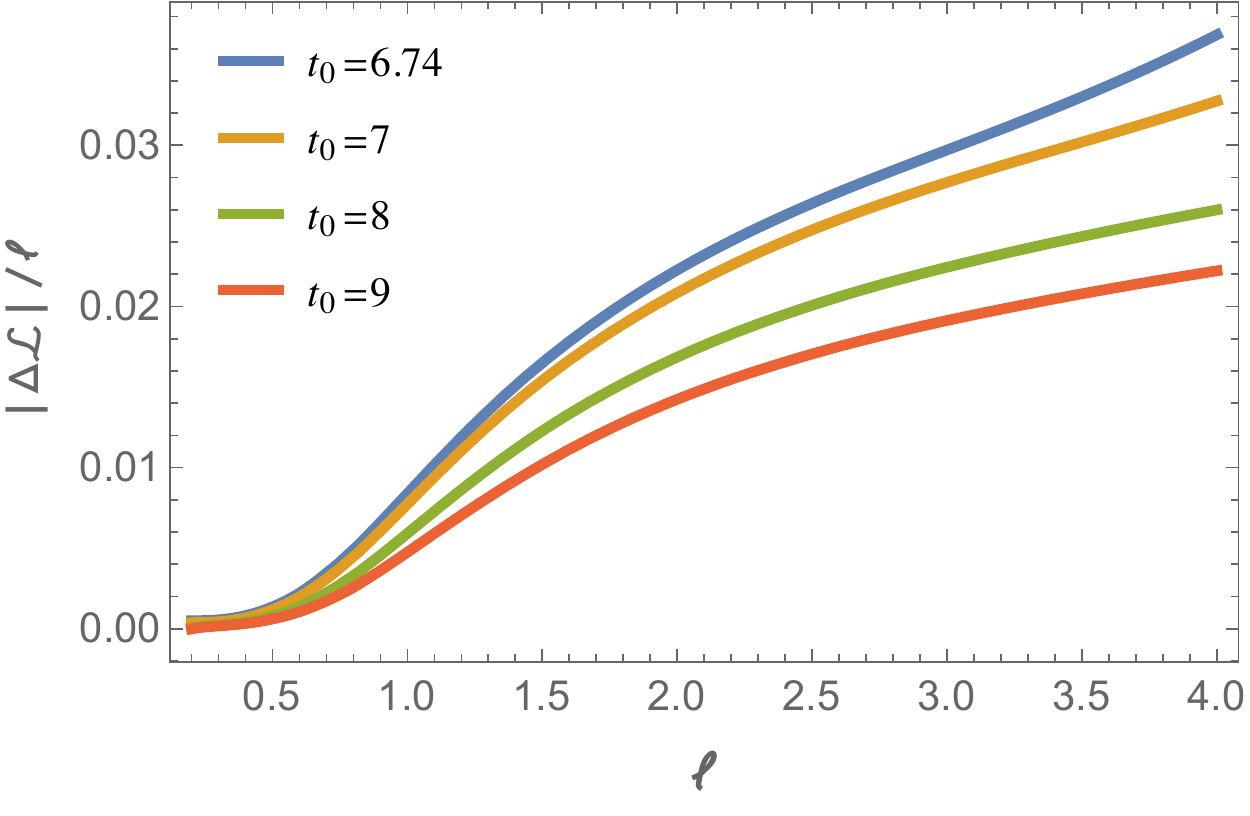}
\caption{\baselineskip 10pt Quench model $\mathcal{B}$: $|\Delta\mathcal{L}|/\ell$ versus $\ell$ at values of $t_0$ specified in the legenda.}\label{fig:flex}
\end{center}
\end{figure}
\item $\ell_{b}$ is the value of $\ell$ where $|\Delta \mathcal{L}|/\ell_{b}= 0.01$, $|\Delta \mathcal{A}_R|/\ell_{b}=0.01$ and $|\Delta \mathcal{A}_C|/\ell_{b}^2= 0.01$. One considers as thermalized the geodesics having $\ell\leqslant \ell_{b}$, where the difference with respect to the hydrodynamic result is less than the chosen bound.
At $t_0=6.74$, we find $\ell_{b}= 1.1$, $\ell_{b}= 0.7$ and $\ell_{b}= 1.6$ for the three observables, respectively.
The values of $\ell_b$ obtained for different $t_0$ are shown in Fig.~\ref{fig:bound}.
\begin{figure}[t!]
\begin{center}
\includegraphics[width = .45\textwidth]{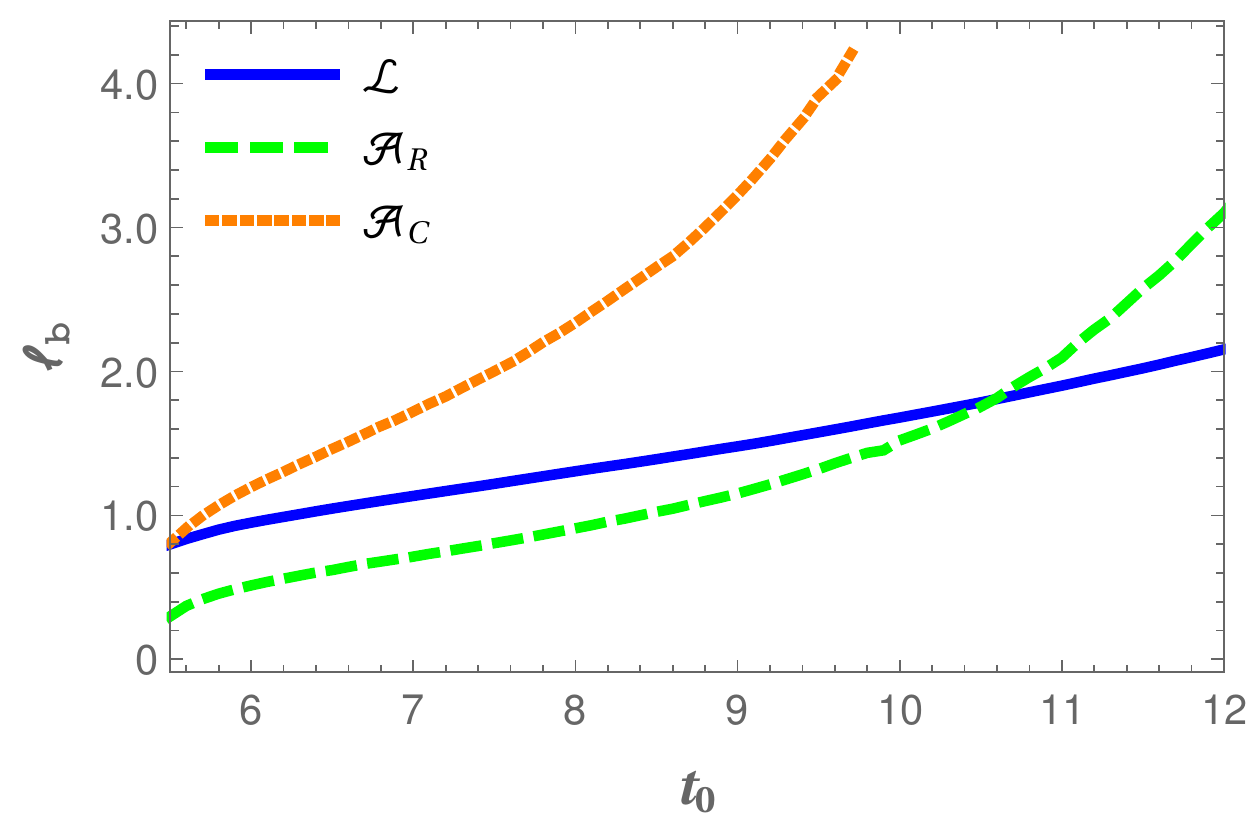}\\
\includegraphics[width = .45\textwidth]{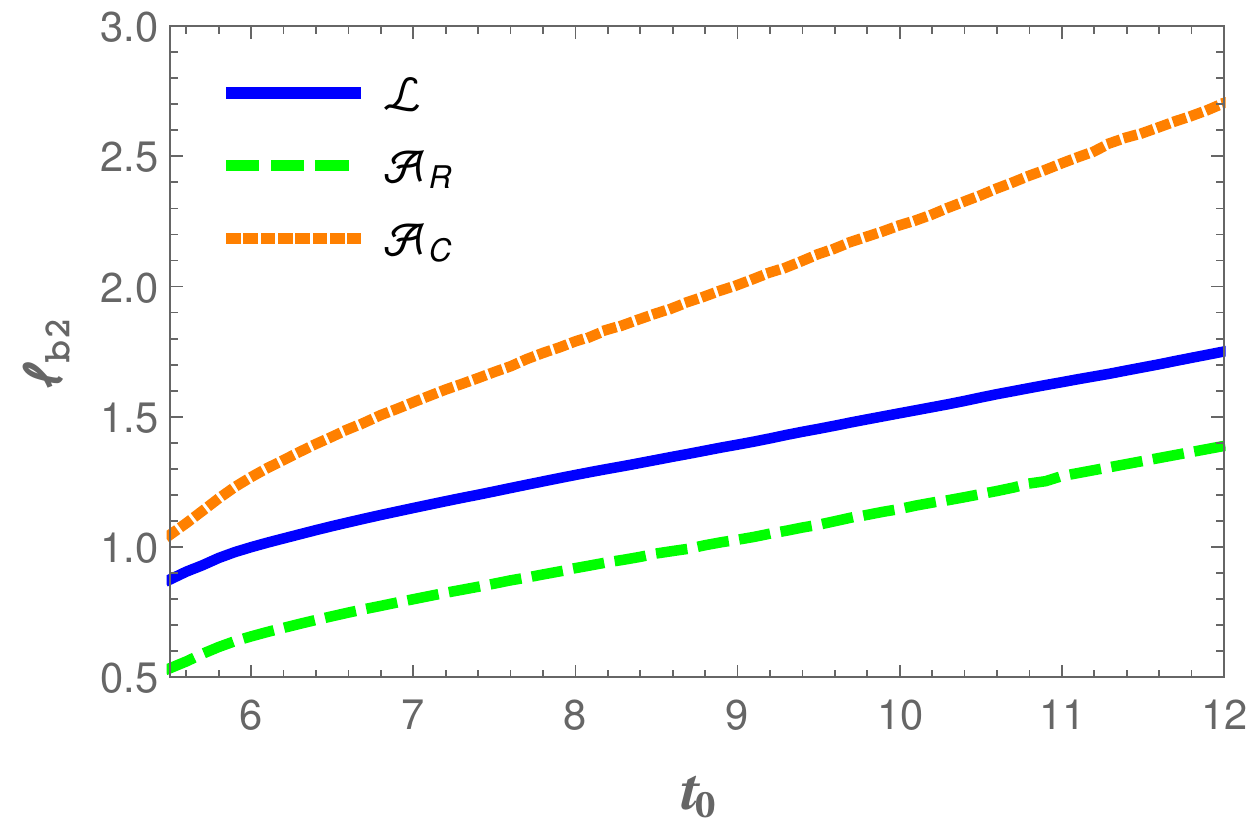}
\caption{\baselineskip 10pt Quench model $\mathcal{B}$: critical sizes $\ell_b$ (top) and $\ell_{b2}$ (bottom) versus $t_0$ for the three nonlocal observables.}\label{fig:bound}
\end{center}
\end{figure}
\item $\ell_{b2}$ is the value of $\ell$ at which $|\Delta \mathcal{L}|/\mathcal{L}= 0.002$, $|\Delta \mathcal{A}_R|/\mathcal{A}_R=0.002$ and $|\Delta \mathcal{A}_C|/\mathcal{A}_C= 0.002$. In this case, one considers thermalized the geodesics having $\ell\leqslant \ell_{b2}$, considering the chosen bound.
At $t_0=6.74$ we find $\ell_{b2}= 1.1$, $\ell_{b2}= 0.8$ and $\ell_{b2}= 1.5$ for the three observables, respectively.
 The values of $\ell_{b2}$ obtained for different $t_0$ are also shown in Fig.~\ref{fig:bound}.
\end{itemize}
The various critical sizes, collected in Table \ref{tab:res}, are consistent with each other, and the  criteria produce a coherent quantitative determination of the thermalization size for the three nonlocal observables.

\begin{figure}[b!]
\begin{center}
\includegraphics[width = .45\textwidth]{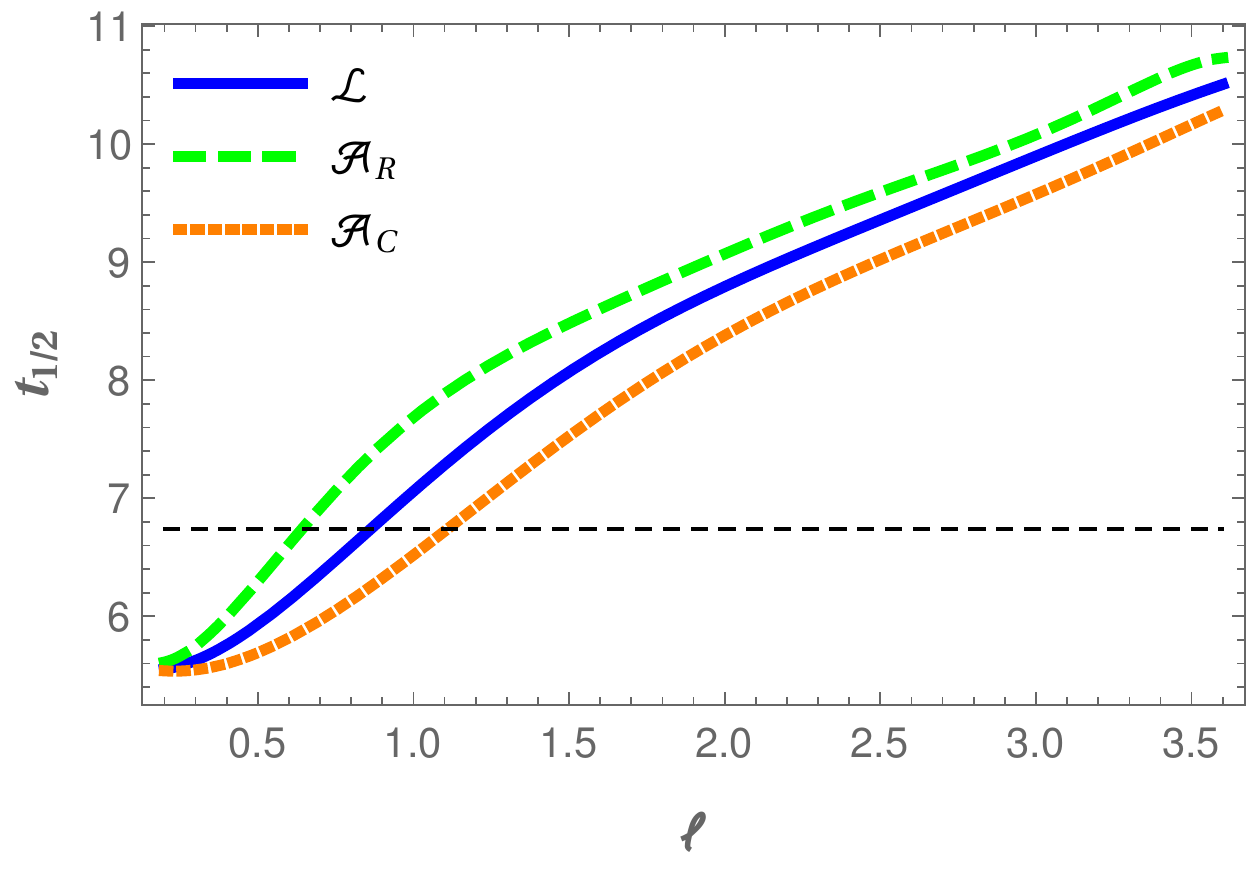}
\includegraphics[width = .45\textwidth]{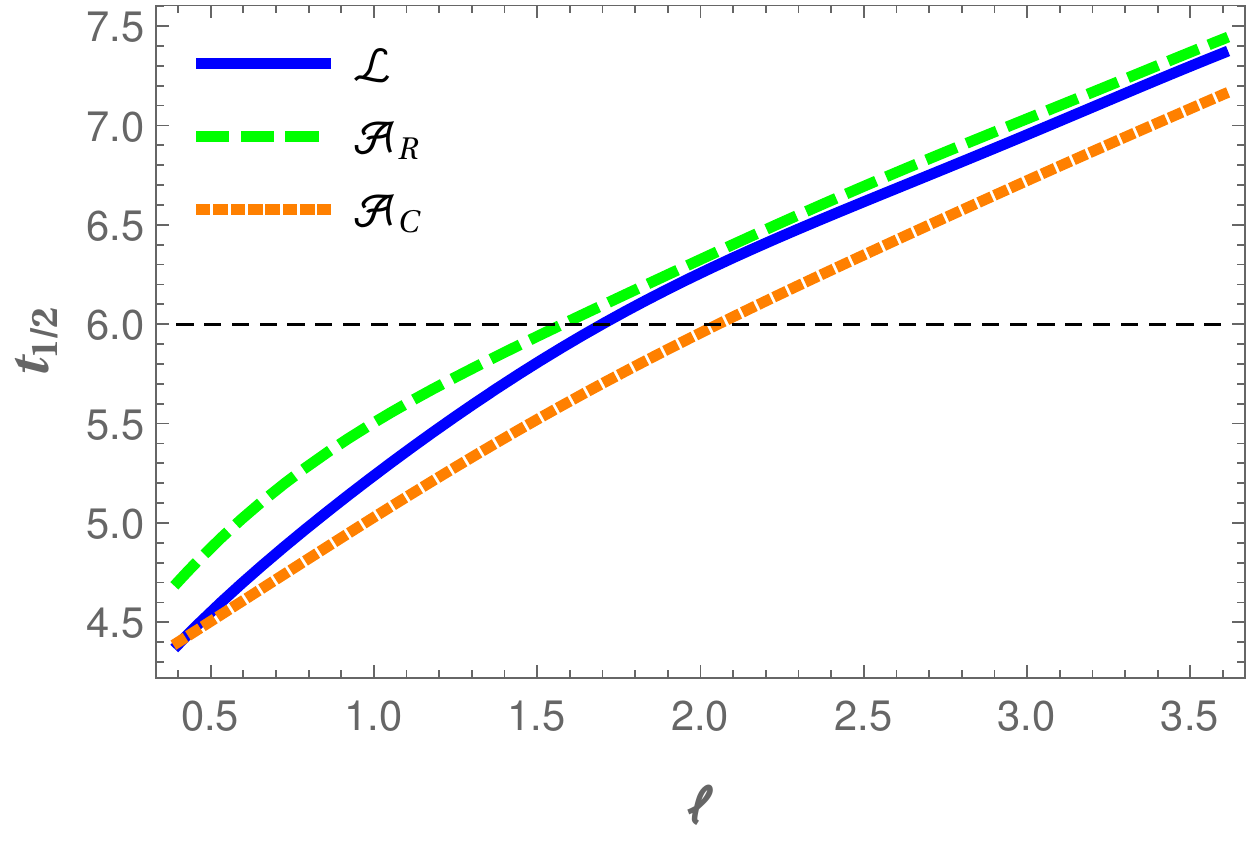}
\caption{\baselineskip 10pt Quench models $\mathcal{B}$ (left) and $\mathcal{A}(2)$ (right): time $t_{1/2}$ versus the size $\ell$ for the three nonlocal observables. The horizontal dashed line indicates the thermalization time obtained from  the pressure anisotropy. }\label{fig:thalf}
\end{center}
\end{figure}

Finally, we define $t_{1/2}(\ell)$ as the value of $t_0$ at which $|\Delta \mathcal{L}|$ is reduced by a half with respect to the end of the quench at fixed $\ell$, and similarly for $|\Delta \mathcal{A}_R|$ and $|\Delta \mathcal{A}_C|$.
The end of the quench is the time $t_0$ corresponding to $\tau_*=5$, a condition ensuring that the whole geodesic $(r(x),\tau(x))$ is not affected by the quench.
The results in Fig.~\ref{fig:thalf} show that $t_{1/2}(\ell)$ exceeds the thermalization time obtained using local observables for size $\ell \simeq 1$, comparable to the critical sizes previously defined.
Regardless of the choice of the criterion, the rectangular Wilson loop takes more time to thermalize.
Another feature emerging for $t_{1/2}$ is the linear increase against the size $\ell$.
This dependence is common to the result obtained in other systems in which the thermalization time for large probes has been scrutinized \cite{Buchel:2014gta}.
The hierarchy found between the thermalization times of the energy density, the pressures and the large-size probes indicates the onset of thermalization starting at short distances.

\begin{table}
\begin{center}
\begin{tabular}{|c|ccc|| ccc |}
\hline
critical size &GL&RWL&CW & GL& RWL& CWL \\ \hline
$\hat \ell_1$&1&0.5&1 & 0.9&0.4&0.8\\
$\hat \ell_2$&0.7&0.3&0.7&0.6&0.3&0.6\\
$\hat \ell_{b}$&1.1&0.7&1.6&0.9&0.5&1.2\\
$\hat \ell_{b2}$&1.1&0.8&1.5 & 1.0& 0.6&1.1 \\ \hline
\end{tabular}
\caption{\baselineskip 10pt Critical sizes for the regularized geodesic length (GL) and the regularized area of rectangular (RWL) and circular Wilson loop (CWL), for quench model $\mathcal{B}$ with $t_0=6.74$ (left) and quench model $\mathcal{A}(2)$ with $t_0=6$ (right). The definitions are given in the text.}\label{tab:res}
\end{center}
\end{table}

\subsection{Quench model $\mathcal{A}(2)$}

We have computed the regularized geodesic lengths and the regularized areas of the extremal surfaces for rectangular and circular Wilson loops in the quench model $\mathcal{A}(2)$ investigated in \cite{Bellantuono:2015hxa}.
The results in Fig.~\ref{fig:lgeo} show how the nonlocal observables follow the quench in the boundary, and how thermalization is reached with the curves approaching the hydrodynamic behavior, as understood by inspecting Fig.~\ref{fig:modB} (right panel).
The curves in the latter figure start at the different  $t_0$ corresponding to $\tau_*=\tau_f^\mathcal{A}=3.25$.
The relaxation to thermalization can be quantified using the same criteria adopted for model $\mathcal{B}$.
The thermalization time found using local observables is $t_0=6$ \cite{Bellantuono:2015hxa}.
From the criterion of the inflection point of $\Delta\mathcal{L}/\ell$, $\Delta\mathcal{A}_R/\ell$ and $\Delta\mathcal{A}_C/\ell^2$  the critical thermalization sizes for the geodesics, the rectangular and the circular Wilson loop are $\hat \ell_1 = 0.9$, $\hat \ell_1 = 0.4$ and $\hat \ell_1 = 0.8$, respectively.
On the other hand, from the inflection point of the derivative of $\Delta\mathcal{L}$, $\Delta\mathcal{A}_R$ and $\Delta\mathcal{A}_C/\ell$, we find $\hat \ell_2 = 0.6$, $\hat \ell_2 = 0.3$ and $\hat \ell_2 = 0.6$, respectively.
The critical sizes $\hat \ell_{b}$ and $\hat \ell_{b2}$, obtained using the same requirements imposed for model $\mathcal B$,
are collected in Table \ref{tab:res} together with $\hat \ell_1$ and $\hat \ell_2$.
The various critical sizes are close to each other, with the different criteria providing a coherent quantitative determination of thermalization for nonlocal observables also in this quench model.

The calculation of the half thermalization time $t_{1/2}(\ell)$ gives the result in Fig.~\ref{fig:thalf}, and the behavior is linear for large sizes. As in model $\mathcal B$, the rectangular Wilson loop thermalizes more slowly than the other two observables.

Let us conclude this section by observing that  in both the quench models we have found the emergence of time scaling related to the onset of hydrodynamics and to the size of the probes. The inspection of the coefficients of the time dependence of, e.g.,  $\Delta \cal L$ in Fig.~\ref{fig:C4C6}  does not allow one to identify other time regimes. This is at odds with the results based on the bulk Vaidya geometry \cite{Liu:2013qca} in which various kinds of time dependences 
are detected for the geodesic length,  related to the so-called pre- and post-equilibrium regimes.  As a last remark, the observation of a hierarchy in thermalization among the different sizes and distances is connected to the use of space-like probes. Analyses based on time correlators, or on horizon-to-boundary propagators in the same dynamical framework, would be useful to show a hierarchy in thermalization among different frequencies and modes of the boundary field theory \cite{CaronHuot:2011dr}.

\section{Conclusions}
In a fully dynamical holographic 5$d$ setup with boundary sourcing, we have studied three nonlocal observables, the equal-time two-point correlation function of a large dimension operator in the boundary theory, and the expectation value of a rectangular and circular Wilson loop.
We have computed the  observables during the quench, for two different models for the distortion of the boundary metric, and after the end of the last spike in the distortion.
The hydrodynamic behavior of the observables has also been determined using a 5$d$ metric reproducing the viscous hydrodynamic time dependence of the energy density and of the transverse and longitudinal pressures.
Thermalization of the nonlocal observables has been scrutinized using the difference between the observables in the quenched and in the hydrodynamic geometries.

In this time-dependent setup, the energy density follows the viscous behavior immediately after the end of the quench, while there is a time delay for the pressures to reach the viscous dependence and the isotropy condition $p_\perp=p_\parallel$  \cite{Bellantuono:2015hxa}. 
For nonlocal observables we have found that the thermalization time changes with the size of the observable. Different criteria defining critical sizes  produce  coherent results.
For larger sizes, the thermalization time increases with the size of the probe.
In particular, for all the three observables the time $t_{1/2}(\ell)$ increases linearly with  $\ell$, a result  independent of the quench model.
The hierarchy among the thermalization times of the energy density, pressures and large probes supports the conclusion of  a faster thermalization  at short distances,  a feature of the strongly coupled theories.

\vspace*{1cm} {\bf \noindent Acknowledgments.}
LB is grateful to Prof. I. Arefeva for interesting discussions.
This work has been carried out within the INFN project (Iniziativa Specifica) QFT-HEP.

\bibliographystyle{apsrev4-1}
\bibliography{FFLPS}
\end{document}